\documentclass[twocolumn]{autart}

\usepackage{amsmath} 
\usepackage{amssymb}
\usepackage{nicefrac}
\usepackage{epstopdf}
\usepackage{graphicx}     
\usepackage{hhline}
\usepackage{amsbsy}
\usepackage{bm}
\usepackage{soul}
\usepackage{units}
\usepackage{siunitx}
\usepackage{mathtools}
\usepackage{mathrsfs}
\usepackage{natbib}
\usepackage{enumerate}
\usepackage{enumitem}
\usepackage{pgfplots}
\usepackage{filecontents}
\usepackage{tikz}
\usepackage{enumerate}
\mathtoolsset{showonlyrefs}
\usepgfplotslibrary{statistics}
\usepackage{pgfplotstable}
\usetikzlibrary{calc}
\usetikzlibrary{patterns}
\usetikzlibrary{positioning}
\usepackage{tikzscale}
\usepackage{booktabs}
\newlength\figureheight
\newlength\figurewidth
\newcommand{\abs}[1]{\left\lvert#1\right\rvert}
\newcommand{\norm}[1]{\left\lVert#1\right\rVert}
\newcommand{\nul}{\text{o}}
\newcommand{\wpone}{\text{ as } N\rightarrow\infty, \text{ w.p.1}}

\newtheorem{assumption}{Assumption}
\newtheorem{definition}{Definition}
\newtheorem{theorem}{Theorem}
\newtheorem{algorithmm}{Algorithm}

\newcommand{\thetals}{\hat{\theta}^\text{LS}_N}
\newcommand{\thetawls}{\hat{\theta}^\text{WLS}_N}

\newcommand{\minus}{\scalebox{0.5}[1.0]{$-$}}
\pgfplotsset{every tick label/.append style={font=\scriptsize}}
\mathtoolsset{showonlyrefs}

\begin{document}

\begin{frontmatter}

\title{Estimating Models with High-Order Noise Dynamics\\Using Semi-Parametric Weighted Null-Space Fitting} 

\thanks[footnoteinfo]{This work was supported by the Swedish Research Council with the project ``System identification: Unleashing the algorithms,'' contract 2015-05285, and the research environment ``NewLEADS---New Directions in Learning Dynamical Systems,`` contract 2016-06079.}

\author[KTH]{Miguel Galrinho}\ead{galrinho@kth.se},    
\author[KTH]{Cristian R. Rojas}\ead{crros@kth.se},               
\author[KTH]{H{\aa}kan Hjalmarsson}\ead{hjalmars@kth.se}  

\address[KTH]{Automatic Control Department, School of Electrical
    Engineering and Data Science,\\
    KTH Royal Institute of Technology, SE-100 44 Stockholm, Sweden.}  


\begin{abstract}                          
Standard system identification methods often provide inconsistent estimates with closed-loop data.
With the prediction error method (PEM), this issue is solved by using a noise model that is flexible enough to capture the noise spectrum.
However, a too flexible noise model (i.e., too many parameters) increases the model complexity, which can cause additional numerical problems for PEM.
In this paper, we consider the weighted null-space fitting (WNSF) method.
With this method, the system is first modeled using a non-parametric ARX model, which is then reduced to a parametric model of interest using weighted least squares.
In the reduction step, a parametric noise model does not need to be estimated if it is not of interest.
Because the flexibility of the noise model is increased with the sample size, this will still provide consistent estimates in closed loop and asymptotically efficient estimates in open loop.
In this paper, we prove these results, and we derive the asymptotic covariance for the estimation error obtained in closed loop, which is optimal for an infinite-order noise model. 
For this purpose, we also derive a new technical result for geometric variance analysis, instrumental to our end.
Finally, we perform a simulation study to illustrate the benefits of the method when the noise model cannot be parametrized by a low-order model.
\end{abstract}

\end{frontmatter}

\section{Introduction}

The prediction error method (PEM) is a benchmark for estimation of linear parametric models.
If the model orders are correct and the noise is Gaussian, PEM with a quadratic cost function is asymptotically efficient~\citep{ljung99}: the asymptotic covariance of the estimates coincides with the Cram{\'e}r-Rao (CR) bound, the lowest covariance attainable by a consistent estimator.

Two models can typically be distinguished in a parametric model structure: the dynamic model and the noise model.
Because the noise sequence is often the result of different noise contributions aggregated in a complex manner, the concept of a ``correct order'' for the noise model is often intractable in practice.
While PEM is still consistent in open loop when the noise model cannot capture the actual noise spectrum, this is not the case for data collected in closed loop. 

This issue with closed-loop data is not exclusive of PEM.
Instrumental variable methods~\citep{ivbook_soderstrom_stoica} require the reference signal in order to construct the instruments in closed loop~\citep{cliv,clriv}.
Classical subspace methods~\citep{vanodem94} also suffer from inconsistency in closed loop, although this issue has been overcome by more recent algorithms: for example, \citet{verhaegen:closedloop} estimates the complete open-loop system followed by a model reduction step; \citet{qin2003closed} estimate the innovations; \citet{jansson03} uses a non-parametric ARX model to construct the Hankel matrices; \citet{chiuso2005consistency} use a whitening filter.

With PEM, the inconsistency issue can in theory be solved by letting the noise model structure be arbitrarily flexible (i.e., letting the number of estimated parameters become arbitrarily large), guaranteeing that a correct noise spectrum can be captured by the model.
If the global minimum of the PEM cost function is found, in open loop this will asymptotically not affect the statistical properties of the dynamic-model estimate; in closed loop, consistency is attained but not efficiency.
The problem with this approach is that, because the noise model might require many parameters, the optimization problem to solve becomes computationally heavier, and the PEM cost function may have more local minima, thus aggravating the numerical search for the global minimum.

Some methods use a semi-parametric approach, estimating a non-parametric model in a first step, which is then used in a second step to estimate the dynamic model.
In this case, an estimate of the noise model is no longer needed for consistency of the dynamic model.
This type of approach is more user-friendly, because the user is not required to choose an appropriate parametric noise model: this has been highlighted by~\citet{freqdom_nonparH}, who propose a frequency-domain method using non-parametric noise models, based on the theory developed by~\citet{schoukens2009nonparametric,Pintelon2010a,Pintelon2010b}.
In the time domain, methods based on the same idea have also been proposed by~\citet{zhu_book,bjsm,MORSM}.
The two latter ones have the advantage of not requiring numerical search algorithms, but they have only been considered for open loop data.

The weighted null-space fitting (WNSF) method~\citep{galrinho14} also first estimates a non-parametric model, and then reduces it to a parametric model, which may or may not include the noise model, and can be used with closed-loop data.
Moreover, similarly to the methods by~\citet{bjsm,MORSM}, WNSF does not apply non-linear optimization techniques, but uses weighted least squares.
In this sense, the method can be seen as belonging to a family of iterative least-squares methods with an intermediate non-parametric model estimate.

Precursors of the method have used an impulse response estimate to obtain a rational transfer function~\citep{evansfischl73,shaw94,shawmisra94,lemmerlingettal01}.
The origin of this family of methods can be traced back to the field of time-series analysis.
\citet{durbin1960fitting} proposes two methods for auto-regressive moving-average (ARMA) time series estimation, using a high-order AR time series as intermediate step to obtain the ARMA parameters by least squares.
The first method is non-iterative but does not attain the CR bound, while the second method remedies this by iterating between estimating the AR and MA polynomials, initialized with the first method.
An alternative way to attain the CR bound from Durbin's first method as starting point, and which simplified the analysis, was proposed by~\citet{mayne1982linear}, using an additional filtering step.

An important theoretical challenge is how to establish consistency and asymptotic efficiency for this type of methods.
The aforementioned papers for estimation of time series consider the non-parametric model order as tending to infinity, but ``small'' compared to the sample size.
In practice, however, it is intuitive that the non-parametric model order is chosen depending on the available sample size. 
In this sense, the theoretical analysis should consider this relation formally, with the non-parametric model order tending to infinity as function of the sample size, according to some specified rate.
This dependency is introduced by~\citet{hannanKavalieris} to prove consistency of the method by~\citet{mayne1982linear}.
Later applications include vector ARMA time-series~\citep{hannan1984multivariate,reinsel1992maximum,dufour2014asymptotic}, for which also asymptotic efficiency has been treated formally.

\citet{wnsf_tac} have derived the asymptotic properties of WNSF for a fully parametric model  (i.e., dynamic and noise models), making use of the results by~\citet{ljung&wahlberg92} on the statistical properties of non-parametric least-squares estimates when the model order tends to infinity as function of the sample size at particular rates.
In particular, the fully parametric WNSF is shown to be consistent and asymptotically efficient with open- and closed-loop data when the correct model structure is chosen.
In this paper, instead of estimating both the dynamic and noise models, we disregard the parametric noise model, reducing the non-parametric model estimate to obtain a parametric dynamic model only. 
The dynamic model estimate will then be asymptotically efficient in open loop, and consistent in closed loop, with optimal asymptotic covariance for an infinite-order noise model. 
The asymptotic properties of the proposed method correspond to the asymptotic properties of PEM with an infinite-order noise model~\citep{Forssell_cl} in both open and closed loop, but performed with a robust numerical procedure, and without using a numerical search algorithm.

The case addressed in this paper, where WNSF is used with no parametric noise-model estimate, will be denoted semi-parametric WNSF.
Despite having been mentioned by~\citet{wnsf_tac}, a formal analysis of the semi-parametric case has not been considered.
Technically, the theoretical analysis of this setting is significantly more challenging than the fully parametric case.
The reason is that expressions with structure $(T R^{-1} T^\prime)^{-1}$, where $T$ and $R$ are matrices, appear in both variants of the method; however, $T$ and $R$ are both square and invertible in the fully parametric case, and the analysis can be done using $T^{-\prime} R T^{-1}$, while this is not the case for the semi-parametric case.
This type of structure arises in the analysis by~\citet{geometric_jonas}, whose geometric approach to variance analysis can be helpful here. There, however, the sizes of the matrices are constant, whereas here the dimensions grow unlimited, causing  important technical issues that need to be resolved. 
This motivates that, in Section \ref{subsec:geomanalysisresult}, we derive a new result based on the approach by~\citet{geometric_jonas} such that it can be applied in the setting of our problem.


The paper is organized in the following way.
In Section~\ref{sec:prel}, we introduce definitions, assumptions, and background.
In Section~\ref{sec:method}, we present the semi-parametric algorithm of WNSF.
In Section~\ref{sec:theoretical}, we provide the asymptotic analysis, which includes: 1) a new result for variance analysis that is instrumental for our problem; 2) consistency with open- and closed-loop data; 3) asymptotic efficiency with open-loop data and the corresponding asymptotic covariance matrix with closed-loop data.
In Section~\ref{sec:sim}, we perform an experimental analysis illustrating the potential of the method.

\section{Preliminaries}
\label{sec:prel}

Most of the notation, definitions, and assumptions used by~\cite{wnsf_tac} apply to this paper.
For completeness, we present them here.

\subsection{Notation}
\label{sec:prel:notation}
\begin{itemize}[noitemsep,nolistsep]
\item $A^\prime$ is the transpose of matrix $A$.
\item $A^*$ is the complex conjugate transpose of matrix $A$.
\item $\norm{x}=\sqrt{\sum_{k=1}^{n}\abs{x_k}^2}$, with $x$ an $n\times 1$ vector.
\item $\norm{A}=\sup_{x\neq 0} \norm{Ax}/\norm{x}$, with $A$ a matrix and $x$ a vector of appropriate dimensions.
\item $||A||_F=\sqrt{\text{Trace}(AA^*)}$ (i.e., the Frobenius norm), with $A$ a matrix.
\item $\norm{\bar{G}(q)}_{\mathcal{H}_2} := \sqrt{\! \frac{1}{2\pi} \int_{-\pi}^{\pi} \text{Trace} \bar{G}(e^{i\omega}) \bar{G}^*(e^{i\omega}) d\omega}$, with $\bar{G}(q)$ a transfer matrix.
\item $\norm{\bar{G}(q)}_{\mathcal{H}_\infty} := \sup_\omega ||\bar{G}(e^{i\omega})||$.
\item $C$ denotes any constant, which need not be the same in different expressions.
\item $\Gamma_n(q) = [q^{-1} \quad \cdots \quad q^{-n}]^\prime$, where $q^{-1}$ is the backward time-shift operator.
\item $\mathcal{T}_{n,m}(X(q))$ is the Toeplitz matrix of size $n\times m$ ($m\leq n$) with first column $[x_0 \; \cdots \; x_{n-1}]^\prime$, where $X(q)=\sum_{k=0}^\infty x_k q^{\- k}$, and zeros above the main diagonal.
\item $\mathbb{E} x$ denotes expectation of the random vector $x$.
\item $\bar{\mathbb{E}} x_t := \lim\limits_{N\to\infty} \frac{1}{N} \sum_{t=1}^{N} \mathbb{E} x_t$.
\item $x_N = \mathcal{O}(f_N)$ means that the function $x_N$ tends to zero at a rate not slower than $f_N$, as $N\to\infty$, w.p.1.
\item $\langle X(q),Y(q) \rangle \!\!:=\!\! \frac{1}{2\pi} \int_{-\pi}^{\pi} \!\!X(e^{i\omega}) Y^*(e^{i\omega}) d\omega$, with $X(q)$ and $Y(q)$ transfer matrices of appropriate sizes.
\end{itemize}

\subsection{Assumptions}

\begin{assumption}[True system and parametric model]
\label{ass:truesystem}
The system has scalar input $\{u_t\}$, scalar output $\{y_t\}$, and is subject to scalar noise $\{e_t\}$. These signals are related by
\begin{equation}
y_t = G_\nul(q)u_t + H_\nul(q) e_t,
\label{eq:truesys}
\end{equation}
where $G_\nul(q)$ and $H_\nul(q)$ are rational functions given by
\begin{equation}
\begin{aligned}
G_\nul(q) &= \frac{L_\nul(q)}{F_\nul(q)} = \frac{l_1^\nul q^{-1} + \cdots + l_{m_l}^\nul q^{-m_l} }{1+ f_1^\nul q^{-1} + \cdots + f_{m_f}^\nul q^{-m_f}},
\\
H_\nul(q) &= \frac{C_\nul(q)}{D_\nul(q)} = \frac{1 + c_1^\nul q^{-1} + \cdots + c_{m_c}^\nul q^{-m_c} }{1+ d_1^\nul q^{-1} + \cdots + d_{m_d}^\nul q^{-m_d}}.
\end{aligned}
\label{eq:GH}
\end{equation}
The transfer functions $G_\nul$, $H_\nul$, and $H_\nul^{-1}$ are assumed to be stable. 
The polynomials $L_\nul$ and $F_\nul$---as well as $C_\nul$ and $D_\nul$---do not share common factors.

We parametrize $G(q)$ as
\begin{equation}
G(q,\theta) = \frac{L(q,\theta)}{F(q,\theta)} = \frac{l_1 q^{-1} + \cdots + l_{m_l} q^{-m_l} }{1+ f_1 q^{-1} + \cdots + f_{m_f} q^{-m_f}},
\label{eq:G}
\end{equation}
where
\begin{equation}
\theta =
\begin{bmatrix}
f_1 & \cdots f_{m_f} & l_1 & \cdots & l_{m_l}
\end{bmatrix}^\prime 
\end{equation}
is the parameter vector, with known orders $m_f$ and $m_l$.
We assume that there is $\theta_\nul$ such that $G(q,\theta_\nul)=G_\nul(q)$.
The orders $m_c$ and $m_d$ of the noise model numerator and denominator polynomials are not known.
\end{assumption}

Because we allow for data to be collected in closed loop, the input $\{u_t\}$ is allowed to have a stochastic part.
Then, let $\mathcal{F}_{t-1}$ be the $\sigma$-algebra generated by ${\{e_s, u_s, s\leq t-1\}}$. 
For the noise, the following assumption applies.

\begin{assumption}[Noise]
\label{ass:noise}
The noise sequence $\{e_t\}$ is a stochastic process that satisfies
\begin{equation}
\mathbb{E}[e_t|\mathcal{F}_{t-1}] = 0, \quad  \mathbb{E}[e_t^2|\mathcal{F}_{t-1}] = \sigma_\nul^2, \quad \mathbb{E}[|e_t|^{10}] \leq C, \forall t .
\end{equation}
\end{assumption}

Before stating the assumption on the input, we introduce some definitions from \cite{ljung&wahlberg92}.

\begin{definition}[$f_N$-quasi-stationarity]
Let $f_N$ be a decreasing sequence of positive scalars, with $f_N \to 0$ as $N \to \infty$, and
\begin{equation}
R^N_{vv}(\tau) = \left \{ 
\begin{array}{lr}
\frac{1}{N}\sum_{t = \tau+1}^N v_tv_{t-\tau}^\prime, & 0 \le \tau < N, \\
\frac{1}{N}\sum_{t = 1}^{N+\tau} v_tv_{t-\tau}^\prime, & -N < \tau \leq 0,\\
0, & \mathrm{otherwise.}
\end{array}
\right.
\end{equation}
The vector sequence $\{v_t\}$ is $f_N$-quasi-stationary if
\begin{enumerate}
\item There exists $R_{vv}(\tau)$ such that \\ $\sup_{|\tau| \leq N} \norm{R^N_{vv}(\tau)- R_{vv}(\tau)} \leq C_1 f_N$,
\item $\frac{1}{N}\sum_{t = -N}^N \norm{v_t}^2  \le C_2$
\end{enumerate}
for all $N$ large enough, where $C_1$ and $C_2$ are finite constants.
\end{definition}

\begin{definition}[$f_N$-stability]
\hspace{-1mm}A filter \hspace{0.2mm} $G(q) \! = \! \sum_{k=0}^\infty  g_k q^{\minus k}$
is $f_N$-stable if $\sum_{k=0}^\infty |g_k| /f_k < \infty $.
\end{definition}

\begin{definition}[Power spectral density]
The power spectral density of an $f_N$-quasi-stationary sequence $\{v_t\}$ is given by $\Phi_{v}(z)=\sum_{\tau=\minus\infty}^\infty R_{vv}(\tau)z^{-\tau}$, if the sum exists for $|z|=1$.
\end{definition}

\begin{assumption}[Input]
\label{ass:input}
The input sequence $\{u_t\}$ is defined by
$u_t = - K(q) y_t + r_t$
under the following conditions.
\begin{enumerate}
\item The sequence $\{r_t\}$ is independent of $\{e_t\}$, $f_N$-quasi-stationary with $f_N\!=\!\sqrt{\log N/N}$, and uniformly bounded.
\item With $\Phi_{r}(z) = F_r(z)F_r(z^{\minus 1})$ the spectral factorization of $\{r_t\}$ and $F_r(z)$ causal, $F_r(q)$ is BIBO stable.
\item The closed loop system is $f_N$-stable with $f_N = 1/\sqrt{N}$.
\item The transfer function $K(z)$ is bounded on the unit circle.
\item The spectral density of $\{[ r_t \; e_t ]^\prime \}$ is bounded from below by the matrix $\delta I$, for some $\delta>0$.
\end{enumerate}
\end{assumption}
Operation in open loop is obtained by taking $K(q)=0$. 

Alternatively to~\eqref{eq:truesys}, the true system can be written as
\begin{equation}
A_\nul(q) y_t = B_\nul(q) u_t + e_t ,
\label{eq:truearx}
\end{equation}
where
\begin{equation}
\begin{alignedat}{3}
A_\nul(q) &:= \frac{1}{H_\nul(q)} &&=: 1+&&\sum_{k=1}^{\infty} a^\nul_k q^{\- k} , \\
B_\nul(q) &:= \frac{G_\nul(q)}{H_\nul(q)} &&=: &&\sum_{k=1}^{\infty} b^\nul_k q^{\- k} 
\end{alignedat}
\label{eq:truearxpoly}
\end{equation}
are stable (Assumption~\ref{ass:truesystem}).
In a first step, WNSF estimates truncated versions of $A_\nul(q)$ and $B_\nul(q)$, using the ARX model
\begin{equation}
A(q,\eta^n) y_t = B(q,\eta^n) u_t + e_t ,
\label{eq:arxmodel}
\end{equation}
where
\begin{gather}
\eta^n = 
\begin{bmatrix}
a_1 & \cdots & a_n & b_1 & \cdots & b_n
\end{bmatrix}^\prime , \label{eq:eta}\\
A(q,\eta^n) = 1+\sum_{k=1}^{n} a_k q^{-k} , \quad B(q,\eta^n) = \sum_{k=1}^{n} b_k q^{-k} .
\end{gather}
Because the order needs to be infinite for the system to be in the ARX model set, we make the model order $n$ depend on the sample size $N$---denoted ${n=n(N)}$---according to the following assumption.

\begin{assumption}[ARX-model order]
\label{ass:ARXorder}
The ARX model order is selected according to:
\begin{itemize}
\item [D1.] $n(N)\to\infty$, as $N\to\infty$;
\item [D2.] $n^{4+\delta}(N)/N \to 0$, for some $\delta>0$, as $N\to\infty$.
\end{itemize}
\end{assumption}

Compared with~\cite{wnsf_tac}, the only difference in the assumptions is on the model: therein also the noise model uses a parametric structure analogous to~\eqref{eq:G}, whereas here it is not estimated.

\subsection{Prediction Error Method}
The prediction error method minimizes a cost function of the prediction errors
\begin{equation}
\varepsilon_t(\theta,\zeta) = H^{-1}(q,\zeta) \left( y_t - \frac{L(q,\theta)}{F(q,\theta)} u_t \right) ,
\label{eq:epsilon}
\end{equation}
where $H(q,\zeta)$ is a noise model, parametrized by
\begin{equation}
H(q,\zeta) = \frac{C(q,\zeta)}{D(q,\zeta)} = \frac{1 + c_1 q^{-1} + \cdots + c_{m_c} q^{-m_c} }{1+ d_1 q^{-1} + \cdots + d_{m_d} q^{-m_d}},
\end{equation}
with
\begin{equation}
\zeta = 
\begin{bmatrix}
c_1 & \cdots & c_{m_c} & d_1 & \cdots & d_{m_d}
\end{bmatrix}^\prime
\end{equation}
With a quadratic cost function, the PEM estimate of the parameters is obtained by minimizing
\begin{equation}
J(\theta,\zeta) = \frac{1}{N} \sum_{t=1}^{N} \frac{1}{2} \varepsilon_t^2(\theta,\zeta) ,
\label{eq:J}
\end{equation}
where $N$ is the sample size.
Using the quadratic cost function~\eqref{eq:J} can provide asymptotically optimal estimates when the noise is Gaussian.

Let $H(q,\zeta)$ be such that there exists $\zeta=\zeta_\nul$ for which $H(q,\zeta_\nul)=H_\nul(q)$.
Denoting by $\hat{\theta}_N^\text{PEM}$ the parameter vector $\theta$ that (together with some $\zeta$) minimizes~\eqref{eq:J}, the estimate $\hat{\theta}_N^\text{PEM}$ is asymptotically distributed as~\citep{ljung99}
\begin{equation}
\sqrt{N} (\hat{\theta}^\text{PEM}_N-\theta_\nul) \sim As\mathcal{N}(0, \sigma_\nul^2 M_\text{PEM}^{-1}) ,
\label{eq:PEMcov}
\end{equation}
where $\mathcal{N}$ stands for the Gaussian distribution. 
Let $\Phi_u^r$ be the spectrum of
\begin{equation}
u_t^r = S_\nul(q)r_t,
\label{eq:utr}
\end{equation}
with $S_\nul(q) = [1+K(q)G_\nul(q)]^{-1}$ the sensitivity function, and $\Gamma_{m_f}$ and $\Gamma_{m_l}$ be according to the definition of $\Gamma_n$ in Section~\ref{sec:prel:notation} with $n=m_f$ and $n=m_l$, respectively.
Then, the asymptotic covariance matrix in~\eqref{eq:PEMcov} satisfies
\begin{equation}
M_\text{PEM} \geq \frac{1}{2\pi} \int_{-\pi}^{\pi} \bar{\Omega}(e^{i\omega}) \Phi_u^r(e^{i\omega}) \bar{\Omega}^*(e^{i\omega}) d\omega =: M ,
\label{eq:M}
\end{equation}
where 
\begin{equation}
\bar{\Omega}(e^{i\omega}) = 
\begin{bmatrix}
-\frac{G_\nul}{H_\nul F_\nul} \Gamma_{m_f} \\
\frac{1}{H_\nul F_\nul} \Gamma_{m_l}
\end{bmatrix} .
\end{equation}
Inside the matrix, arguments were omitted for notational simplicity; however, for clarity, we point out that functions of $q$ are evaluated at $q=e^{i\omega}$ when inside of integrals that have the frequency $\omega$ as variable of integration.

In open loop (in which case $\Phi_u^r$ is simply the input spectrum), ${M_\text{PEM}=M}$, and it corresponds to the CR bound under a Gaussian noise assumption.
In closed loop, $M_\text{PEM}=M$ when the number of parameters in $\zeta$ tends to infinity.
In this case, $M$ does not correspond to the CR bound, but to the optimal covariance (from a prediction error perspective) with an infinite-order noise model~\citep{Forssell_cl}.
For additional discussion on open- and closed-loop accuracy, we refer to~\citet{bombois2011olvscl,aguero2007olvscl}

The interest of estimating a non-parametric noise model in closed loop is that even if the noise spectrum needs to be captured by a high-order model, it will still be possible to obtain a consistent estimate of the dynamic model $G(q,\theta)$.
However, estimating a non-parametric noise model simultaneously with a parametric dynamic model with PEM is not realistic.
The reason is that, as the number of parameters in $H(q,\zeta)$ increases, the prediction error~\eqref{eq:epsilon} becomes a more complicated function of $\zeta$, which makes the problem computationally heavier, and more difficult to find the global minimum of the non-convex cost function~\eqref{eq:J}.
Consequently, the theoretically attractive result that PEM with a non-parametric noise model provides estimates with covariance corresponding to $M$ may not always be useful in practice.
It turns out that this setting can be handled with WNSF  without increasing the difficulty of the problem.

\section{Semi-Parametric Weighted Null-Space Fitting Algorithm}
\label{sec:method}

The WNSF method consists of three steps~\citep{wnsf_tac}.
First, we estimate a non-parametric ARX model, with least squares.
Second, we reduce this estimate to a parametric model, with least squares, providing a consistent estimate.
Third, we re-estimate the parametric model, with weighted least squares, where an estimate of the optimal weighting is used to attain an asymptotically efficient estimate.
We now consider the procedure for each step, without estimating a parametric noise model.
Because WNSF has been already presented by~\citet{galrinho14,wnsf_tac}, we will not go into detail here on the motivation and derivation of each equation, but instead focus on the idea of the semi-parametric algorithm.

For the first step, consider~\eqref{eq:arxmodel} in the regression form
\begin{gather}
y_t = (\varphi_t^n)^\prime \eta^n + e_t , \\
\varphi_t^n =
\begin{bmatrix}
-y_{t-1} & \cdots & - y_{t-n} & u_{t-1} & \cdots & u_{t-n}
\end{bmatrix}^\prime.
\end{gather}
Then, the least-squares estimate of $\eta^n$ is obtained by
\begin{equation}
\hat{\eta}^{n}_N = [R^n_N]^{-1} r^n_N ,
\label{eq:ls}
\end{equation}
where
\begin{equation}
R^n_N = \frac{1}{N} \sum_{t=n+1}^N \varphi_t^n (\varphi_t^n)^\prime , \quad
r^n_N = \frac{1}{N} \sum_{t=n+1}^N \varphi_t^n y_t ,
\label{eq:RnN}
\end{equation}
for which we have that~\citep{ljung&wahlberg92} 
\begin{equation}
\begin{alignedat}{3}
R^n_N &\to \bar{R}^n&& :=\bar{\mathbb{E}}\left[ \varphi_t^n (\varphi_t^n)^\prime \right] , &&\wpone , \\
r^n_N &\to \bar{r}^n&& :=\bar{\mathbb{E}}\left[ \varphi_t^n y_t \right] , &&\wpone ,\\
\hat{\eta}^{n}_N &\to \bar{\eta}^n&& := \left[\bar{R}^n\right]^{-1} \bar{r}^n, &&\wpone .
\end{alignedat}
\label{eq:Rbar}
\end{equation}

For the second step, we obtain an estimate of $G(q,\theta)$, from the non-parametric ARX-model estimate.
For this purpose, we may use~\eqref{eq:GH} and~\eqref{eq:truearxpoly} to write
\begin{equation}
\begin{aligned}
C_\nul(q)A_\nul(q) - D_\nul(q) &= 0  , \\
F_\nul(q)B_\nul(q) - L_\nul(q)A_\nul(q) &= 0 .
\end{aligned}
\label{eq:CA-D=0andFB-LA=0} 
\end{equation}
Because we are not interested in estimating a parametric noise model, the first equation in~\eqref{eq:CA-D=0andFB-LA=0} is not relevant for our purposes.
Then, we require only
\begin{equation}
F_\nul(q)B_\nul(q) - L_\nul(q)A_\nul(q) = 0.
\label{eq:FB-LA=0}
\end{equation}
By convolution, \eqref{eq:FB-LA=0} can be written in matrix form as
\begin{equation}
b^n_\nul - Q_n(\eta^n_\nul) \theta_\nul = 0 ,
\label{eq:eta-Qtheta=0}
\end{equation}
where $\eta^n_\nul$ is given by~\eqref{eq:eta} evaluated at the true coefficients of~\eqref{eq:truearxpoly}, $b^n_\nul$ consists of 
$b^n =
[b_1 \; \dots \; b_n]^\prime$ also evaluated at the true coefficients, and
\begin{gather}
Q_n(\eta^n) =
\begin{bmatrix}
-Q^f_n(\eta^n) & Q^l_n(\eta^n)
\end{bmatrix} , \\
Q^l_n(\eta^n) = \mathcal{T}_{n,m_l}(A(q,\eta^n)), \;
Q^f_n(\eta^n) = \mathcal{T}_{n,m_f}(B(q,\eta^n)),
\label{eq:Q}
\end{gather}
where $\mathcal{T}_{n\times m_f}$ is according to the definition in Section~\ref{sec:prel:notation} with $m=m_f$.
Motivated by~\eqref{eq:eta-Qtheta=0}, we replace $\eta^n_\nul$ by its estimate $\hat{\eta}^{n}_N$ (and the same for $b^n_\nul$, which is a part of $\eta^n_\nul$) and obtain an estimate of $\theta$ with least squares:
\begin{equation}
\hat{\theta}_N^\text{LS} = \left( Q_n^\prime(\hat{\eta}^{n}_N) Q_n(\hat{\eta}^{n}_N) \right)^{-1} Q_n^\prime(\hat{\eta}^{n}_N) \hat{b}^{n}_N .
\label{eq:theta_ls}
\end{equation}
This estimate, as will be shown in Theorem~\ref{thm:consistencyLS}, is consistent.

For the third step, we re-estimate $\theta$ to obtain an asymptotically efficient estimate.
This is done by taking into account the statistical properties of the errors in $\hat{\eta}^{n}_N$.
As $\eta^n_\nul$ is replaced by $\hat{\eta}^{n}_N$ in~\eqref{eq:eta-Qtheta=0}, the residuals can be written as
\begin{equation}
\hat{b}^{n}_N - Q_n(\hat{\eta}^{n}_N)\theta_\nul = T_n(\theta_\nul) (\hat{\eta}^{n}_N - \eta^n_\nul) ,
\label{eq:residualsWNSF}
\end{equation}
where
\begin{gather}
T_n(\theta) = 
\begin{bmatrix}
-T^l_n(\theta) & T^f_n(\theta)
\end{bmatrix}, \\
T^l_n(\theta) = \mathcal{T}_{n,n}(L(q,\theta)) , \;
T^f_n(\theta) = \mathcal{T}_{n,n}(F(q,\theta)) .
\label{eq:T0}
\end{gather}
For the term $\hat{\eta}^{n}_N - \eta^n_\nul$ in~\eqref{eq:residualsWNSF}, if we neglect the bias error originating from the truncation taking place in the ARX model (which should be close to zero for sufficiently large $n$) we have that, approximately,
\begin{equation}
\sqrt{N} \left(\hat{\eta}^{n}_N-\bar{\eta}^n\right) \sim As\mathcal{N} (0,\sigma_\nul^2 [\bar{R}^n]^{-1} ).
\label{eq:eta_dist}
\end{equation}
This allows us to express the covariance of the residuals~\eqref{eq:residualsWNSF} as being proportional to
\begin{equation}
\bar{W}_n^{-1}(\theta_\nul) = T_n(\theta_\nul) [\bar{R}^n]^{-1} T^\prime_n(\theta_\nul),
\label{eq:barWninv}
\end{equation}
Using the inverse  of~\eqref{eq:barWninv} as weighting, when solving~\eqref{eq:residualsWNSF} in a least-squares sense, minimizes the variance of the parameter estimate.
Although this covariance is dependent on the true parameters, a consistent estimate is available from Step 2.
Hence, we may use as weighting
\begin{equation}
W_n(\hat{\theta}^\text{LS}_N) = \Big(T_n(\hat{\theta}^\text{LS}_N) [R^n]^{-1} T^\prime_n(\hat{\theta}^\text{LS}_N)\Big)^{-1}.
\label{eq:Wninv}
\end{equation}
and the estimate obtained in this step is thus given by
\begin{equation}
\hat{\theta}^{\text{WLS}}_N \!\!= \Big( Q_n^\prime(\hat{\eta}^{n}_N) W_n(\hat{\theta}^{\text{LS}}_N) Q_n(\hat{\eta}^{n}_N) \Big)^{-1} \!\!\! Q_n^\prime(\hat{\eta}^{n}_N) W_n(\hat{\theta}^{\text{LS}}_N) \hat{b}^{n}_N .
\label{eq:theta_wls}
\end{equation}

The algorithm may be summarized as follows.
\begin{algorithmm}
\label{alg:spWNSF}
The semi-parametric WNSF method consists of the following steps:
\begin{enumerate}
\item compute a non-parametric ARX-model estimate with~\eqref{eq:ls};
\item compute a parametric dynamic-model estimate with~\eqref{eq:theta_ls};
\item re-compute a parametric dynamic-model estimate with~\eqref{eq:theta_wls}.
\end{enumerate}
\end{algorithmm}

Optionally, we may continue to iterate, potentially improving the estimation quality for finite sample size.
However, we show in the next section that Algorithm~\ref{alg:spWNSF} has the same asymptotic properties as PEM with an infinite-order noise model.
Nevertheless, the algorithm has advantages with respect to PEM with an arbitrarily flexible noise model.
First, WNSF estimates the non-parametric noise model in a separate step, as part of an ARX model, which is linear in the model parameters; thus, it does not make the problem computationally more difficult, unlike if~\eqref{eq:J} is minimized with an arbitrary large number of noise model parameters.
Second, it overall does not use numerical search algorithms that can converge to non-global optima.
The price to pay is that, even if the noise spectrum can be modeled parametrically, WNSF still requires a noise model whose order tends to infinity in order to satisfy the statistical properties that we proceed to show.

\section{Theoretical Analysis}
\label{sec:theoretical}

In this section, we perform a theoretical analysis of the semi-parametric WNSF method.
In particular, we will show that Step 3 in Algorithm 1 provides a consistent estimate and derive its covariance matrix.
For that, we will need some auxiliary results.

\subsection{Results from~\citet{wnsf_tac}}

To show the aforementioned results, we will need that the estimate obtained in Step 2 of Algorithm~\ref{alg:spWNSF} is consistent.
For that, we have the following result.
\begin{theorem}
\label{thm:consistencyLS}
Let Assumptions~\ref{ass:truesystem}, \ref{ass:noise}, \ref{ass:input}, and~\ref{ass:ARXorder} hold, and let $\hat{\theta}_N^\text{LS}$ be given by~\eqref{eq:theta_ls}.
Then,
\begin{equation}
\hat{\theta}_N^\text{LS} \to \theta_\nul, \wpone .
\end{equation}
\end{theorem}

\begin{pf}
For the fully parametric case, where $\theta$ additionally contains the noise model parameters, the analogous result is shown by~\citet{wnsf_tac} in Theorem 1.
Therein, the idea of the proof is to consider separately the part of the expression in~\eqref{eq:theta_ls} providing the noise model estimates and the dynamic model estimates, as the two problems are separable when no weighting is used.
The part corresponding to the dynamic model, which in turn corresponds to $Q(\eta^n)$ given by~\eqref{eq:Q}, are identical for both fully- and semi-parametric cases.
Hence, the result follows from Theorem 1 in~\citet{wnsf_tac}.
\hfill\hfill\qed
\end{pf}

Although consistency of Step 2 with semi-parametric WNSF is a specific case of the results for the fully-parametric method, consistency and asymptotic covariance of Step 3 are technically more challenging to derive.
In the following subsection, we provide further insight into why, and derive a result that will be instrumental for the remainder of our analysis.

\subsection{Result for variance analysis with geometric approach}
\label{subsec:geomanalysisresult}
We begin by writing, for the estimate from Step 3,
\begin{multline}
\thetawls-\theta_\nul \\ = M^{-1}(\hat{\eta}_N,\thetals) Q_n^\prime(\hat{\eta}_N) W_n(\thetals) T_n(\theta_\nul) (\hat{\eta}_N-\eta^{n(N)}_\nul) ,
\label{eq:thetawls-theta0_expand}
\end{multline}
where we define $M(\eta^n,\theta) := Q_n^\prime(\eta^n) W_n(\theta) Q_n(\eta^n)$ and ${\hat{\eta}_N := \hat{\eta}_N^{n(N)}}$, recalling that $n$ is a function of $N$ according to Assumption~\ref{ass:ARXorder} (for notational simplicity, we use only $n$ in matrix subscripts even when it is a function of $N$).
To analyze the asymptotic properties of semi-parametric WNSF, the limit value of~\eqref{eq:thetawls-theta0_expand} and the asymptotic distribution of $\sqrt{N}(\thetawls-\theta_\nul)$ will be considered.
The technical challenge in this analysis, compared to~\citet{wnsf_tac}, comes from the matrix $W_n(\thetals)$, and consequently also the matrix $M(\hat{\eta}_N,\thetals)$, which contains $W_n(\thetals)$.
Considering~\eqref{eq:T0}, we observe that the outer inverse in~\eqref{eq:Wninv} cannot be computed by taking the individual inverses contained in $W_n(\thetals)$, as consequence of $T_n(\thetals)$ not being square.
On the other hand, in the fully parametric case, the matrix $T_n(\thetals)$ is square and converges to an invertible matrix~\citep{wnsf_tac}. 
Hence, we may write $W_n(\thetals)=T_n^{-\prime}(\thetals) R^n_N T_n^{-1}(\thetals)$ for the fully parametric case, which is used throughout the analysis by~\citet{wnsf_tac}, but not for the semi-parametric case considered in this paper.

To deal with this issue, we use the approach by~\cite{geometric_jonas}, writing the aforementioned matrices as projections of the rows of some matrix onto the subspace spanned by the rows of another matrix.
This will be applied to the limit value of the matrix $M(\hat{\eta}_N,\thetals)$, defined by
\begin{equation}
\bar{M}(\eta_\nul,\theta_\nul) \!\!:=\!\!\! \lim_{n\to\infty} \! Q_n^\prime(\eta^n_\nul) [T_n(\theta_\nul) (\bar{R}^n)^{-1} T^\prime_n(\theta_\nul)]^{-1} Q_n(\eta^n_\nul) . 
\label{eq:Mbar}
\end{equation}
Writing $Q_n(\eta_\nul^n)$, $T_n(\theta_\nul)$, and $\bar{R}^n$ (defined in~\eqref{eq:Q}, \eqref{eq:T0}, and~\eqref{eq:RnN}, respectively) in the frequency domain, we have
\begin{equation}
\begin{aligned}
&Q_n^\prime(\eta^n_\nul) = \frac{1}{2\pi} \int_{-\pi}^{\pi}
\begin{bmatrix}
\Gamma_{m_f} & 0 \\ 0 & \Gamma_{m_l}
\end{bmatrix}
\begin{bmatrix}
-B_\nul \\ A_\nul 
\end{bmatrix}
\Gamma_n^* d\omega, \\
&T_n(\theta_\nul) = \frac{1}{2\pi} \int_{-\pi}^{\pi}
\begin{bmatrix}
\Gamma_n & \Gamma_n
\end{bmatrix}
\begin{bmatrix}
-L_\nul^* & 0 \\ 0 & F^*_\nul 
\end{bmatrix}
\begin{bmatrix}
\Gamma_n^* & 0 \\ 0 & \Gamma_n^*
\end{bmatrix}
d\omega , \\
&\begin{multlined}[\displaywidth]
\bar{R}^n = \frac{1}{2\pi} \int_{-\pi}^{\pi}
\begin{bmatrix}
\Gamma_n & 0 \\ 0 & \Gamma_n
\end{bmatrix}
\begin{bmatrix}
-G_\nul S_\nul F_r & -H_\nul S_\nul \sigma_\nul \\ S_\nul F_r & -K H_\nul S_\nul \sigma_\nul
\end{bmatrix} \\ \cdot
\begin{bmatrix}
-G_\nul S_\nul F_r & -H_\nul S_\nul \sigma_\nul \\ S_\nul F_r & -K H_\nul S_\nul \sigma_\nul
\end{bmatrix}^*
\begin{bmatrix}
\Gamma_n^* & 0 \\ 0 & \Gamma_n^*
\end{bmatrix}
d\omega.
\end{multlined}
\end{aligned}
\end{equation}
Arguments are omitted for notational simplicity, but functions of $q$ in the time domain should be evaluated at $q=e^{i\omega}$.
Using these expressions, we may write $\bar{M}(\eta_\nul,\theta_\nul)$ as
\begin{multline}
\bar{M}(\eta_\nul,\theta_\nul) = \\ \lim_{n\to\infty} \langle\gamma,{\Psi_n}\rangle [ \langle{\Psi_n},{\Omega_n}\rangle \langle{\Omega_n},{\Omega_n}\rangle^{-1} \langle{\Omega_n},{\Psi_n}\rangle ]^{-1} \langle{\Psi_n},\gamma\rangle,
\label{eq:Mbar_proj}
\end{multline}
where 
\begin{equation}
\begin{aligned}
{\Omega_n} &=
\begin{bmatrix}
-G_\nul(q)S_\nul(q) \Gamma_n(q) & -\sigma_\nul H_\nul(q) S_\nul(q) \Gamma_n(q) \\
S_\nul(q)F_r(q) \Gamma_n(q) & -\sigma_\nul K(q) H_\nul(q) S_\nul(q) \Gamma_n(q)
\end{bmatrix} , \\
{\Psi_n} &=
\frac{F_\nul^*(q)}{S_\nul^*(q)F_r^*(q)}
\begin{bmatrix}
\Gamma_n(q) & 0
\end{bmatrix} , \\
\gamma &=
\frac{S_\nul(q)F_r(q)}{F_\nul(q)}
\begin{bmatrix}
-B_\nul(q) \Gamma_{m_f}(q) & 0 \\ 
A_\nul(q) \Gamma_{m_l}(q) & 0
\end{bmatrix} .
\label{eq:gammaPsiOmega_original}
\end{aligned}
\end{equation}
Following the approach by~\cite{geometric_jonas}, we recognize that the term $$\langle{\Psi_n},{\Omega_n}\rangle \langle{\Omega_n},{\Omega_n}\rangle^{-1} \langle{\Omega_n},{\Psi_n}\rangle$$ in~\eqref{eq:Mbar_proj} can be written as
\begin{equation}
\langle{\Psi_n},{\Omega_n}\rangle \langle{\Omega_n},{\Omega_n}\rangle^{-1} \langle{\Omega_n},{\Psi_n}\rangle = \langle \text{Proj}_{\mathcal{S}_{\Omega_n}}{\!\!\!\Psi_n}, \text{Proj}_{\mathcal{S}_{\Omega_n}}{\!\!\!\Psi_n} \rangle ,
\end{equation}
where $\text{Proj}_{\mathcal{S}_{\Omega_n}}{\Psi_n}$ denotes the projection of the rows of ${\Psi_n}$ onto the subspace spanned by the rows of ${\Omega_n}$.
As $n\to\infty$, the dimensions of the matrix $\Omega_n$ increase, and the subspace spanned by its rows approaches $\mathcal{H}_2$.
Then, the limit value of the projection will be the causal part of the projected matrix.

For a simplified case, suppose that $\Psi_n$ were causal and that its dimension did not depend on $n$ (i.e., $\Psi_n=\Psi$).
In this case, we would have $\lim_{n\to\infty} \langle \text{Proj}_{\mathcal{S}_{\Omega_n}}{\Psi}, \text{Proj}_{\mathcal{S}_{\Omega_n}}{\Psi} \rangle=\langle\Psi,\Psi\rangle$.
In turn, we would then have that $\bar{M}(\eta_\nul,\theta_\nul) = \langle \gamma,\Psi \rangle \langle \Psi,\Psi \rangle \langle \Psi, \gamma \rangle  = \langle \text{Proj}_{\mathcal{S}_{\Psi}}{\gamma}, \text{Proj}_{\mathcal{S}_{\Psi}}{\gamma} \rangle$.
If we now reintroduce that the dimension of $\Psi$ depends on $n$ ($\Psi=\Psi_n$), and we assume that the rows of $\Psi_n$ span $\mathcal{H}_2$ as $n\to\infty$, we have that $\bar{M}(\eta_\nul,\theta_\nul) =\lim_{n\to\infty}\langle \text{Proj}_{\mathcal{S}_{\Psi_n}}{\gamma}, \text{Proj}_{\mathcal{S}_{\Psi_n}}{\gamma} \rangle = \langle \gamma,\gamma \rangle$.
These arguments follow from results in~\cite{geometric_jonas}. 
However, handling the dimensional increase of $\Psi_n$ with $n$ requires additional technical developments.
One of the key results for the asymptotic analysis in this paper is that the aforementioned result (i.e., that $\bar{M}(\eta_\nul,\theta_\nul) = \langle \gamma,\gamma \rangle$) still holds when the dimensions of $\Psi_n$ increase with $n$.
This is considered in the following theorem.

\begin{theorem}
\label{thm:geometric}
Let
\begin{gather}
{\Omega_n} =
\begin{bmatrix}
F_1(e^{i\omega}) \Gamma_n(e^{i\omega}) & F_2(e^{i\omega}) \Gamma_n(e^{i\omega}) \\
F_3(e^{i\omega}) \Gamma_n(e^{i\omega}) & F_2(e^{i\omega}) F_4(e^{i\omega}) \Gamma_n(e^{i\omega})
\end{bmatrix} , \\
{\Psi_n} =
\begin{bmatrix}
F_5^*(e^{i\omega}) \Gamma_n(e^{i\omega}) & 0
\end{bmatrix} , \;\;
\gamma =
\begin{bmatrix}
F_6(e^{i\omega}) \Gamma_{m_f}(e^{i\omega}) & 0 \\ 
F_7(e^{i\omega}) \Gamma_{m_l}(e^{i\omega}) & 0
\end{bmatrix} ,
\label{eq:gammaPsiOmega}
\end{gather}
where $F_j(q) = \sum_{k=0}^\infty f^{(j)}_k q^{-k}$ ($j=\{1,...,6\}$) are exponentially stable (i.e., ${|f^{(j)}_k|<C\lambda^k\text{ }\forall j}$, $\lambda<1$), $f^{(4)}_0\neq 0$, and $[F_3(q) - F_1(q)F_4(q)]^{-1}$ and $F_5^{-1}(q)$ are exponentially stable.
Then, if there is $\bar{n}$ such that $\sigma_\text{\emph{min}}(\langle{\Omega_n},{\Omega_n}\rangle)<C$ ($\sigma_\text{\emph{min}}$ denotes the smallest singular value) 
for all $n>\bar{n}$,
\begin{multline}
\lim\limits_{n\to\infty} \langle\gamma,{\Psi_n}\rangle [ \langle{\Psi_n},{\Omega_n}\rangle \langle{\Omega_n},{\Omega_n}\rangle^{-1} \langle{\Omega_n},{\Psi_n}\rangle ]^{-1} \langle{\Psi_n},\gamma\rangle \\ = \langle\gamma,\gamma\rangle.
\label{eq:proj}
\end{multline}
\end{theorem}

\begin{pf}
See Appendix~\ref{sec:proof_lemma1}. \hfill\hfill\qed
\end{pf}

\subsection{Consistency and Asymptotic Covariance}

Using the results derived above, we can show the asymptotic properties of semi-parametric WNSF.
Regarding consistency of Step 3 in Algorithm~\ref{alg:spWNSF}, we have the following result.
\begin{theorem}
\label{thm:consistencyWLS}
Let Assumptions~\ref{ass:truesystem}, \ref{ass:noise}, \ref{ass:input}, and~\ref{ass:ARXorder} hold.
Then, 
\begin{equation}
\hat{\theta}_N^\text{WLS} \to \theta_\nul, \wpone .
\end{equation}
\end{theorem}
\begin{pf}
See Appendix~\ref{sec:theorem2_proof}. \hfill\hfill\qed
\end{pf}
\noindent
Theorem~\ref{thm:consistencyWLS} implies that semi-parametric WNSF provides consistent estimates of $\theta_\nul$.

Regarding the asymptotic distribution and covariance of Step 3 in Algorithm~\ref{alg:spWNSF}, we have the following result.
\begin{theorem}
\label{thm:asycov}
Let Assumptions~\ref{ass:truesystem}, \ref{ass:noise}, \ref{ass:input}, and~\ref{ass:ARXorder} hold.
Then, 
\begin{equation}
\sqrt{N}(\hat{\theta}_N^\text{WLS} - \theta_\nul) \sim As\mathcal{N}(0,\sigma_\nul^2 M^{-1}) ,
\label{eq:thetawls_dist}
\end{equation}
where
$M$ is given by~\eqref{eq:M}.
\end{theorem}
\begin{pf}
See Appendix~\ref{sec:theorem3_proof}. 
\end{pf}
\noindent
As consequence of Theorem~\ref{thm:asycov}, the semi-parametric WNSF method summarized in Algorithm~\ref{alg:spWNSF} has the same asymptotic distribution and covariance as PEM with an infinite-order noise model~\citep{ljung99,Forssell_cl}.
In open loop and for Gaussian noise, this corresponds to an asymptotically efficient estimate.

\section{Simulations}
\label{sec:sim}

In this section, we perform three simulation examples.
In the first, we illustrate the asymptotic properties of the method.
In the second, we use a scenario where a low-order parametrization for the noise model does not capture the noise spectrum accurately, which is specially advantageous for semi-parametric WNSF compared with PEM with different choices of noise model.
In the third, we motivate the advantage of semi-parametric WNSF compared to the fully parametric version when using the previous scenario.

\subsection{Illustration of asymptotic properties}

As consequence of the results in Section~\ref{sec:theoretical}, semi-parametric WNSF is asymptotically efficient in open loop for Gaussian noise, with the asymptotic covariance of the dynamic-model estimates given by $\sigma_\nul^2 M^{-1}$.
In closed loop, the asymptotic covariance is still given by $\sigma_\nul^2 M^{-1}$, but in this case it does not correspond to the CR bound, but to the optimal asymptotic covariance when the noise-model order tends to infinity.

To illustrate this, we perform open- and closed-loop simulations such that the closed-loop data are generated by
\begin{equation}
\begin{aligned}
u_t &= \frac{1}{1+K(q)G_\nul(q)} r_t  -  \frac{K(q)H_\nul(q)}{1+K(q)G_\nul(q)} e_t, \\
y_t  &=  \frac{G_\nul(q)}{1+K(q)G_\nul(q)} r_t  + \frac{H_\nul(q)}{1+K(q)G_\nul(q)} e_t, 
\end{aligned}
\label{eq:Kbelow_CL}
\end{equation}
and the open-loop data by
\begin{equation}
\begin{aligned}
u_t &= \frac{1}{1+K(q) G_\nul(q)} r_t  , \\
y_t &= G_\nul(q) u_t  + H_\nul(q) e_t, \\
\end{aligned}
\end{equation}
where $\{r_t\}$ and $\{e_t\}$ are independent Gaussian white sequences with unit variance, $K(q)=1$, and
\begin{equation}
G_\nul(q) = \frac{q^{- 1} + 0.1q^{- 2}}{1 -0.5 q^{- 1} + 0.75 q^{- 2}} , \quad H_\nul(q) = \frac{1 + 0.7 q^{- 1}}{1 - 0.9 q^{- 1}} .
\label{eq:sim_G_H}
\end{equation}

We perform 1000 Monte Carlo runs, with sample sizes $N\in\{300,600,1000,3000,6000,10000\}$.
We apply WNSF with an ARX model of order 50 to both the open- and closed-loop data.
Performance is evaluated by the mean-squared error of the estimated parameter vector of the dynamic model,
\begin{equation}
\text{MSE} = ||\hat{\theta}^\text{WLS}_N-\bar{\theta}_\nul||^2,
\label{eq:nmse}
\end{equation}
As this simulation has the purpose of illustrating asymptotic properties, initial conditions are assumed known---that is, the sums in~\eqref{eq:RnN} start at $t=1$ instead of $t=n+1$.

\begin{figure}
\centering
%
%
\definecolor{mycolor1}{rgb}{0.09804,0.32941,0.65098}%
\definecolor{mycolor3}{rgb}{0.61569,0.06275,0.17647}%
\definecolor{mycolor2}{rgb}{0.38431,0.57255,0.18039}%
\definecolor{mycolor4}{rgb}{0.98039,0.72549,0.09804}%
\definecolor{mycolor5}{rgb}{0.84706,0.32941,0.59216}%
\definecolor{mycolor6}{rgb}{0.39608,0.39608,0.42353}%
\pgfplotsset{compat=1.14}
\begin{tikzpicture}

\begin{axis}[%
width=0.4\textwidth,
height=0.45*0.4\textwidth,
at={(0.758in,0.352in)},
scale only axis,
xmode=log,
xmin=300,
xmax=10000,
xminorticks=false,
ymode=log,
ymin=1e-04,
ymax=0.05,
yminorticks=false,
xlabel=$N$,
ylabel=MSE,
ytick={1e-4,1e-3,1e-2},
axis x line=bottom,
axis y line=left,
x label style={at={(axis description cs:.5,-.12)},anchor=west,font=\footnotesize},
y label style={at={(axis description cs:-.1,.5)},anchor=south,font=\footnotesize},
axis background/.style={fill=white},
legend style={legend cell align=left,align=left,draw=white!15!black}
]
\addplot [color=black,solid]
  table[row sep=crcr]{%
300	0.010304620957736\\
600	0.0042467522071607\\
1000	0.0022885393728325\\
3000	0.000643886274105536\\
6000	0.000320966466169624\\
10000	0.000181210308802515\\
};


\addplot [color=black,dashed]
  table[row sep=crcr]{%
300	0.0205292582688859\\
600	0.00532397012419582\\
1000	0.00260471151656709\\
3000	0.00066287122133017\\
6000	0.000326819065563849\\
10000	0.000184174067184447\\
};



\addplot [color=black,dotted]
  table[row sep=crcr]{%
300	0.006524\\
600	0.003262\\
1000	0.0019572\\
3000	0.0006524\\
6000	0.0003262\\
10000	0.00019572\\
};

\end{axis}
\end{tikzpicture}%
\caption{Illustration of asymptotic properties: theoretical asymptotic MSE (dotted) and average MSE for the parameter estimates as function of sample size obtained with semi-parametric WNSF in closed loop (solid) and open loop (dashed).}
\label{fig:asymp_eff}
\end{figure}
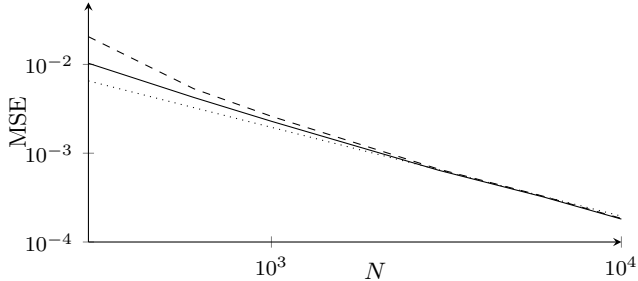

The results are presented in Fig.~\ref{fig:asymp_eff}, with the average MSE plotted as function of the sample size (closed loop in solid line, open loop in dashed line).
We plot also $\sigma_\nul^2 \text{Trace}[M^{-1}]/N$ (dotted line), which the average MSE attains both in open and closed loop: because the data were generated such that $\Phi_u^r$, the spectrum of~\eqref{eq:utr}, is the same for both data sets, both scenarios have the same asymptotic covariance, in accordance to our theoretical results.

\subsection{Random noise model}

When a low-order parametrization of the noise model is not enough to capture the noise spectrum, the noise model may require many parameters.
With PEM, a simultaneous estimate of the dynamic model and a long noise model is not numerically robust due to the non-convexity of the cost function.
The semi-parametric WNSF is appropriate to deal with this scenario, because the noise spectrum is captured beforehand with a non-parametric ARX model.

Modeling the correct noise spectrum is particularly important in closed loop, as the estimates of the dynamic model will be inconsistent if the noise model is not flexible enough to capture the noise spectrum.
For this reason, we consider a closed-loop setting, where data are generated by
\begin{equation}
\begin{aligned}
u_t = \frac{1}{1+K(q)G_\nul(q)} r_t - \frac{K(q)H_\nul(q)}{1+K(q)G_\nul(q)} e_t \\
y_t = \frac{G_\nul(q)}{1+K(q)G_\nul(q)} r_t + \frac{H_\nul(q)}{1+K(q)G_\nul(q)} e_t .
\end{aligned}
\end{equation}
The signals $\{r_t\}$ and $\{e_t\}$ are Gaussian white noise sequences with variances 1 and 4, respectively.
The system is given by
\begin{equation}
G_\nul(q) = \frac{1.0q^{-1}-0.80q^{-2}}{1-0.95q^{-1}+0.90q^{-2}},
\end{equation}
the controller by $K(q)=0.2$, and the true noise model by
\begin{equation}
H_\nul(q) = 1+\sum_{k=1}^{N-1} \lambda_k q^{-k}
\label{eq:H}
\end{equation}
with $\lambda_k = w_k e^{-0.2k}$, where $w_k$ is drawn from a Gaussian distribution with zero mean and unit variance.
Here, differently than Assumption~\ref{ass:truesystem}, stability of $H_\nul(q)$ is not ensured.
However, this is not an issue if the noise is Gaussian, as there always exists an inversely stable $H_\nul(q)$ for which the noise sequence has the same spectrum.

To model the noise, one possibility is to try to find an appropriate low-order parametrization in the form
\begin{equation}
H(q,\zeta) = \frac{c_1 q^{-1} + \cdots + c_{m_h} q^{-{m_h}}}{d_1 q^{-1} + \cdots + d_{m_h} q^{-{m_h}}},
\label{eq:Hzeta}
\end{equation}
where
$\zeta = 
[c_1 \; \dots \; c_{m_h} \; d_1 \; \dots \; d_{m_h}]^\prime.$
In this case, a Box-Jenkins model is estimated.
The most appropriate noise model order may be chosen by using some information criterion, such as the Akaike Information Criterion (AIC) or the Bayesian Information Criterion (BIC)~\citep{ljung99}.

The alternative is to use a high-order model for the noise model. 
For example, 
\begin{equation}
H(q,\zeta^n) = 1+\sum_{k=1}^n \zeta_k q^{-k},
\label{eq:Hzetan_alternative}
\end{equation}
or
\begin{equation}
H(q,\zeta^n) = \frac{1}{1+\sum_{k=1}^n \zeta_k q^{-k}},
\label{eq:Hzetan}
\end{equation}
where
$\zeta^n = [\zeta_1 \; \dots \; \zeta_n]^\prime$.
In particular, the choice~\eqref{eq:Hzetan} has the same structure as the noise model estimated by semi-parametric WNSF.

Motivated by these alternatives, we compare the following methods:
\begin{itemize}
\item semi-parametric WNSF, with non-parametric ARX model order $n=200$ (denoted WNSF$_\text{sp}$);
\item PEM, with default MATLAB initialization, and noise model~\eqref{eq:Hzeta} with $m_h\in\{1,2,...30\}$, where the order $m_h$ is chosen using the AIC or BIC criterion (denoted $\text{PEM}_\text{aic}$ and $\text{PEM}_\text{bic}$, respectively);
\item PEM, with noise model~\eqref{eq:Hzetan} with $n=200$, and MATLAB default initialization (denoted $\text{PEM}_\text{sp}$);
\item PEM, with noise model~\eqref{eq:Hzetan} with $n=200$, and initialized by WNSF$_\text{sp}$ (denoted $\text{PEM}_\text{spWNSF}$);
\end{itemize}
PEM uses the implementation in MATLAB2016b System Identification Toolbox.
All the methods use a maximum of 100 iterations, but stop early upon convergence (default settings for PEM, $10^{-4}$ as tolerance for the normalized relative change in the parameter estimates for WNSF).
The search algorithm used by PEM is chosen automatically.
The noise model~\eqref{eq:Hzetan_alternative} was not used with PEM for computational reasons: the optimization becomes extremely slow as stability of the inverse of the noise model when computing the prediction errors~\eqref{eq:epsilon} is difficult to fulfill with so many parameters, while the inverse of any estimate of~\eqref{eq:Hzetan} is always stable.

We use sample sizes $N\in\{1000,5000,10000\}$ and perform 100 Monte Carlo runs.
Performance is evaluated by the FIT of the impulse response of the dynamic model, given by
\begin{equation}
\text{FIT} = 100\left(1 -\frac{\norm{g_\nul-\hat{g}}}{\norm{g_\nul-\text{mean}(g_\nul)}}\right) ,
\label{eq:fit}
\end{equation}
in percent, where $g_\nul$ is a vector with the impulse response parameters of $G_\nul(q)$ ($\text{mean}(g_\nul)$ is its average), and similarly for $\hat{g}$ but for the estimated model.
In~\eqref{eq:fit}, sufficiently long impulse responses are taken to make sure that the truncation of their tails does not affect the FIT.

\begin{figure}
\centering
\pgfplotsset{width=0.21\textwidth,height=5cm,compat=1.14}
\begin{tikzpicture}
\begin{axis}[
boxplot/draw direction=y,
name = vw_sw,
x axis line style={opacity=0},
axis x line*=bottom,
axis y line=left,
enlarge y limits,
ymin = -31,
xtick style={draw=none},
xticklabels={, , , , , },
legend columns=3,
legend entries={\footnotesize{WNSF$_\text{sp}$},\footnotesize{PEM$_\text{aic}$},\footnotesize{PEM$_\text{bic}$},\footnotesize{PEM$_\text{sp}$},\footnotesize{PEM$_\text{spWNSF}$}},
legend to name=named,
legend style={/tikz/every even column/.append style={column sep=0.5cm}},
ylabel=FIT,
title={$N=1000$},
y label style={at={(axis description cs:-0.15,.45)},anchor=south,font=\footnotesize},]
\addplot [mark=x,boxplot,fill=gray!80,area legend]
table[y index=0] {N1000_wnsf.csv};
\addplot [mark=x,boxplot,pattern=dots,pattern color=gray!80,area legend]
table[y index=0] {N1000_pemaic.csv};
\node at (axis cs:2,-28) [anchor=north] {\scriptsize{$\downarrow\!3$}};
\addplot [mark=x,boxplot,pattern=crosshatch,pattern color=gray!80,area legend]
table[y index=0] {N1000_pembic.csv};
\node at (axis cs:3,-28) [anchor=north] {\scriptsize{$\downarrow\!4$}};
\addplot [mark=x,boxplot,pattern=grid,area legend]
table[y index=0] {N1000_pemn.csv};
\node at (axis cs:4,-28) [anchor=north] {\scriptsize{$\downarrow\!8$}};
\addplot [mark=x,boxplot,area legend]
table[y index=0] {N1000_pemn_wnsf.csv};
\end{axis}

\begin{axis}[
boxplot/draw direction=y,
at = (vw_sw.right of north east),
anchor = north west,
xshift = 2em,
name = vc_sw,
ymin=46,
x axis line style={opacity=0},
axis x line*=bottom,
axis y line=left,
enlarge y limits,
title={$N=5000$},
xtick style={draw=none},
xticklabels={, , , , , }]
\addplot [mark=x,boxplot,fill=gray!80,area legend]
table[y index=0] {N5000_wnsf.csv};
\node at (axis cs:1,47) [anchor=north] {\scriptsize{$\downarrow\!1$}};
\addplot [mark=x,boxplot,pattern=dots,pattern color=gray!80,area legend]
table[y index=0] {N5000_pemaic.csv};
\addplot [mark=x,boxplot,pattern=crosshatch,pattern color=gray!80,area legend]
table[y index=0] {N5000_pembic.csv};
\addplot [mark=x,boxplot,pattern=grid,area legend]
table[y index=0] {N5000_pemn.csv};
\node at (axis cs:4,47) [anchor=north] {\scriptsize{$\downarrow\!4$}};
\addplot [mark=x,boxplot,area legend]
table[y index=0] {N5000_pemn_wnsf.csv};
\end{axis}

\begin{axis}[
boxplot/draw direction=y,
at = (vc_sw.right of north east),
anchor = north west,
name = vc_sc,
xshift = 2em,
ymin  = 46,
x axis line style={opacity=0},
axis x line*=bottom,
axis y line=left,
enlarge y limits,
title={$N=10000$},
xtick style={draw=none},
xticklabels={, , , , , }]
\addplot [mark=x,boxplot,fill=gray!80,area legend]
table[y index=0] {N10000_wnsf.csv};
\addplot [mark=x,boxplot,pattern=dots,pattern color=gray!80,area legend]
table[y index=0] {N10000_pemaic.csv};
\node at (axis cs:2,47) [anchor=north] {\scriptsize{$\downarrow\!4$}};
\addplot [mark=x,boxplot,pattern=crosshatch,pattern color=gray!80,area legend]
table[y index=0] {N10000_pembic.csv};
\node at (axis cs:3,47) [anchor=north] {\scriptsize{$\downarrow\!5$}};
\addplot [mark=x,boxplot,pattern=grid,area legend]
table[y index=0] {N10000_pemn.csv};
\node at (axis cs:4,47) [anchor=north] {\scriptsize{$\downarrow\!4$}};
\addplot [mark=x,boxplot,area legend]
table[y index=0] {N10000_pemn_wnsf.csv};
\end{axis}
\hspace{-3mm}
\node[yshift=-1.8em] at ($(vw_sw.south)!0.5!(vc_sc.south)$) {\pgfplotslegendfromname{named}}; 

\end{tikzpicture}
\vspace{-3mm}
\caption{FITs for given methods and sample sizes from 100 Monte Carlo runs.}
\label{fig:sim}
\end{figure}
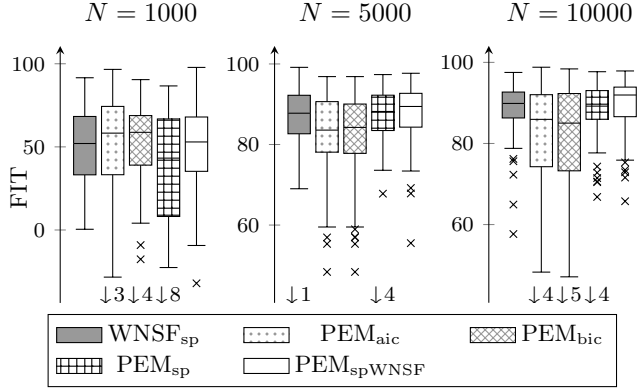

The FITs for the different sample sizes are shown in Fig.~\ref{fig:sim}.
For $N=1000$, semi-parametric WNSF has better performance than PEM with default MATLAB initialization, both with low-order noise model chosen with AIC/BIC and with non-parametric noise model.
Among the PEM alternatives initialized by default with MATLAB, an AIC/BIC criterion with a Box-Jenkins model with orders up to 30 performed better than using a non-parametric noise model.
However, if initialized with semi-parametric WNSF, PEM with non-parametric noise model performs considerably better than with default MATLAB initialization.
Nevertheless the initializing estimate WNSF$_\text{sp}$ may in sometimes be better than the resulting estimate PEM$_\text{spWNSF}$, which can be due to problems with over-fitting.
For $N=5000$, WNSF and PEM with non-parametric noise and default MATLAB initialization have similar median performance, but the algorithm for PEM failed more often.
This can be remedied by initializing PEM with WNSF$_\text{sp}$.
Here, PEM with AIC/BIC had no low-performance outliers, but the median performance was poorer than for the semi-parametric approaches.
Similar conclusions can be drawn for $N=10000$, where PEM does not necessarily provide better results with more data samples, potentially due to numerical problems.

Overall, WNSF showed more robust performance among the sample sizes used than the different variants of PEM with MATLAB initialization.
However, a more evident advantage is the computational time.
Table~\ref{tbl:times} shows the average times, in seconds, required for the identification using semi-parametric WNSF, PEM with all the orders computed for AIC and BIC, $\text{PEM}_\text{sp}$, and $\text{PEM}_\text{spWNSF}$, for the different sample sizes (all the computations were performed in the same computer).
Here, we observe how WNSF requires much lower computational time than the alternatives.
This is a consequence of PEM estimating the noise model in the non-linear optimization procedure, whereas in WNSF the high-order model is estimated in a previous least-squares step, which is numerically robust.
Moreover, the time required for WNSF and PEM with AIC/BIC does not change significantly with $N$, unlike with PEM$_\text{sp}$.
In this case, the time does not necessarily decrease for smaller $N$.
The problem arising when using smaller $N$ is that the cost function more likely becomes ill-conditioned at some parameter values during the optimization.
The time required for PEM with non-parametric noise model decreases when semi-parametric WNSF is used for initialization.
However, the times are still significantly larger than for semi-parametric WNSF, while the improvement in performance in not considerable (Fig.~\ref{fig:sim}).

\begin{table}
  \caption{Average computational times in seconds for WNSF, the search among all orders required for PEM$_\text{aic}$ and PEM$_\text{bic}$, and PEM$_\text{np}$.}
  \label{tbl:times}
  \begin{center}
    \begin{tabular}{ c c c c c }
      $N$                  & WNSF$_\text{sp}$ & $\text{PEM}_\text{aic,bic}$ & $\text{PEM}_\text{np}$ & PEM$_\text{spWNSF}$ \\
      \midrule
      1000 & 1.0  & 30 & 641 & 83 \\
      5000 & 0.91 & 27 &  133 & 42\\
      10000 & 1.3 & 38 & 236 & 82 \\
      \bottomrule
    \end{tabular}
  \end{center}
\end{table}

\subsection{Comparison with fully-parametric WNSF}

In the following, we motivate the advantage of using semi-parametric WNSF compared to the fully-parametric version when the noise cannot be modeled with a low-order parametrization.
In closed loop, and unlike PEM, fully-parametric WNSF should provide consistent estimates of the dynamic model even with an under-parametrized noise model (although this case is not covered in the analysis by~\cite{wnsf_tac}, it should follow from a similar approach).
In this sense, to obtain consistent estimates, fully-parametric WNSF can be used with any choice of noise model.
Nevertheless, this does not render semi-parametric WNSF useless, as we will see with the following illustration.

Consider, besides the WNSF$_\text{sp}$ result from the previous simulation, the following methods:
\begin{itemize}
\item fully-parametric WNSF with noise model~\eqref{eq:Hzeta} and order $m_h=1$ (denoted WNSF$_1$);
\item fully-parametric WNSF with noise model~\eqref{eq:Hzeta} and order $m_h=30$ (denoted WNSF$_{30}$).
\end{itemize}
The number of iterations and stopping criterion is as in the previous simulation, and we use $N=5000$.

\begin{figure}
\centering
\pgfplotsset{width=0.3\textwidth,height=5cm,compat=1.14}
\begin{tikzpicture}
\begin{axis}[
boxplot/draw direction=y,
x axis line style={opacity=0},
axis x line*=bottom,
axis y line=left,
enlarge y limits,
ymin = -40,
xshift = -1cm,
xtick style={draw=none},
xticklabels={, , , , , },
legend columns=-1,
legend entries={\footnotesize{WNSF$_\text{sp}$},\footnotesize{WNSF$_1$},\footnotesize{WNSF$_{30}$}},
legend to name=named,
legend style={/tikz/every even column/.append style={column sep=0.7cm}},
ylabel=FIT,
title={$N=5000$},
y label style={at={(axis description cs:-0.3,.45)},anchor=south,font=\footnotesize},]
\addplot [mark=x,boxplot,fill=gray!80,area legend]
table[y index=0] {N5000_wnsf.csv};
\addplot [mark=x,boxplot,pattern=dots,pattern color=gray!80,area legend]
table[y index=0] {N5000_wnsfH1.csv};
\addplot [mark=x,boxplot,pattern=crosshatch,pattern color=gray!80,area legend]
table[y index=0] {N5000_wnsfH30.csv};
\end{axis}
\node[yshift=-1.8em] {\pgfplotslegendfromname{named}}; 
\end{tikzpicture}
\vspace{-3mm}
\caption{FITs for semi-parametric and fully-parametric WNSF from 100 Monte Carlo runs.}
\label{fig:simWNSFcomp}
\end{figure}
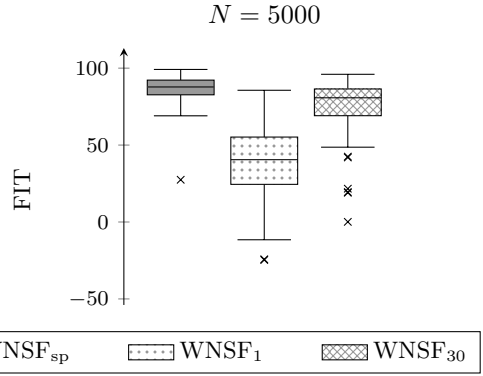

The FIT results are presented in Fig.~\ref{fig:simWNSFcomp}.
The fully-parametric WNSF, both with noise-model order $m_h=1$ and $m_h=30$, performs worse than the semi-parametric version.
In the case that $m_h=1$, this may be due to wrong noise model coefficients being included in the weighting, which, despite not affecting consistency of the dynamic-model estimates, may affect the results for finite sample sizes.
For $m_h=30$, the same reasoning may still apply; in addition, because 30 estimated coefficients of the noise model are included in the weighting, which may have high variance, this may also deteriorate the dynamic-model estimates for finite sample sizes.
These results suggest that when a high-order noise model is required, it may be better to use semi-parametric WNSF than the fully-parametric version.

\section{Conclusion}
Many standard system identification methods provide inconsistent estimates with closed-loop data.
In the particular case of PEM, this issue is avoided by choosing a noise-model order that is large enough to capture the noise spectrum.
An appropriate order is often difficult to choose, and making it arbitrarily large increases the numerical problems of PEM.
The WNSF method, analyzed by~\citet{wnsf_tac} in the fully parametric setting, can also be used without a parametric noise-model estimate, which we named semi-parametric WNSF.
In this paper, we deepened this discussion.

A simulation study illustrates the importance of separating the dynamic- and noise-model identification when a high-order noise model is required, both in terms of performance and computational time.
With WNSF, this separation always occurs, as the method first estimates a non-parametric ARX model.
Then, a parametric noise model does not need to be obtained, as the noise spectrum has been captured in the first step.

We also provide a theoretical analysis of the asymptotic properties.
To this end, we extend the geometric approach by~\citet{geometric_jonas}, deriving a more general result with the matrix dimensions in the inner projection tending to infinity.
We show that semi-parametric WNSF provides consistent estimates of the dynamic model.
With open-loop data, the estimates are also asymptotically efficient;
with closed-loop data, the asymptotic covariance of the estimates corresponds to the best possible covariance with a non-parametric noise model.
This gives WNSF attractive features in terms of flexibility of noise-model structure and asymptotic properties: if a correct parametric noise model is estimated, the dynamic-model estimates are asymptotically efficient; if not, they are consistent and optimal for a non-parametric noise model.
We used a simulation study to illustrate how semi-parametric WNSF is an appropriate method for scenarios where the noise model cannot be accurately modeled with a low-order parametrization.

\appendix
\section{Auxiliary Results for the Proof of Theorem~\ref{thm:geometric}}
The following results will be used to prove Theorem~\ref{thm:geometric}.
\subsubsection*{Cauchy-Schwarz inequality for transfer-matrix inner products}
Let $X(q)$ and $Y(q)$ be transfer matrices and $x$ and $y$ be vectors of appropriate dimensions.
Then, we have
\begin{equation}
\begin{aligned}
||\langle X, Y\rangle||^2 &= \sup_{||x||=1,||y||=1} \abs{ \frac{1}{2\pi} \int_{-\pi}^{\pi} x^\prime X(e^{i\omega}) Y^*(e^{i\omega}) y \text{d}\omega } \\
&\leq \sup_{||x||=1,||y||=1} \sqrt{ x^\prime \langle X,X\rangle x \; y^\prime \langle Y,Y\rangle y } \\
&=||\langle X, X\rangle|| \; ||\langle Y, Y\rangle||. 
\end{aligned}
\label{eq:CS-matrices}
\end{equation}

\subsubsection*{Bound for spectral norm of transfer-matrix inner products}
Let $X(q)$ be a transfer matrix.
Then, we have
\begin{equation}
\begin{aligned}
||\langle X,X \rangle|| &\leq  ||\langle X,X \rangle||_F= \sqrt{\text{Trace} (\langle X,X \rangle^2)} \\
&\leq \text{Trace} \langle X,X \rangle = ||X||_{\mathcal{H}_2}^2,
\end{aligned}
\label{eq:H2bound}
\end{equation}
where the second inequality follows from $\text{Trace}(A^2)\leq[\text{Trace}(A)]^2$ for a positive semi-definite matrix $A$. 

\subsubsection*{Bound for Toeplitz operators of stable filters}

Let $X(q):=\sum_{k=-\infty}^\infty x_k q^{-k}$, with $||X(q)||_{\mathcal{H}_\infty}<C$.
From Theorem 3.1 by~\citet{partington_1989}, it follows that
\begin{equation}
\left|\left| \frac{1}{2\pi} \int_{-\pi}^\pi \Gamma_n \Gamma_n^* X(e^{i\omega}) \right|\right| \leq ||X(q)||_{\mathcal{H}_\infty} \; \forall n\in\mathbb{N}.
\label{eq:ToepBound}
\end{equation}

\section{Proof of Theorem~\ref{thm:geometric}}
\label{sec:proof_lemma1}

In this appendix, we prove Theorem~\ref{thm:geometric}.

\subsection*{Inner Projection: $\langle{\Psi_n},{\Omega_n}\rangle \langle{\Omega_n},{\Omega_n}\rangle^{-1} \langle{\Omega_n},{\Psi_n}\rangle$}

Let
\begin{equation}
{\Psi_n} = {\Psi_n^c} + {\Psi_n^a},
\label{eq:Psic+a}
\end{equation}
where
\begin{equation}
\begin{aligned}
{\Psi_n^c} := 
&\begin{bmatrix}
f^{(5)}_0 e^{-i\omega} & 0 \\ 
f^{(5)}_0 e^{-2i\omega}+f^{(5)}_1 e^{-i\omega} & 0 \\
\vdots & \vdots \\
f^{(5)}_0 e^{-ni\omega} + \sum_{k=1}^{n-1} f^{(5)}_{n-k} e^{-ki\omega} & 0
\end{bmatrix} , \\
{\Psi_n^a} :=
&\begin{bmatrix}
\sum_{k=0}^\infty f^{(5)}_{k+1} e^{i\omega k} & 0 \\ 
\sum_{k=0}^\infty f^{(5)}_{k+2} e^{i\omega k} & 0 \\
\vdots & \vdots \\
\sum_{k=0}^\infty f^{(5)}_{k+n} e^{i\omega k} & 0
\end{bmatrix} .
\end{aligned}
\end{equation}
Alternatively, $\Psi_n^c$ can be written as
${\Psi_n^c} = 
\begin{bmatrix}
P_n \Gamma_n & 0
\end{bmatrix}$,
where $P_n = \mathcal{T}_{n\times n} (F_5(q))$.
Using~\eqref{eq:Psic+a}, we can write
\begin{equation}
\begin{aligned}
&\langle{\Psi_n},{\Omega_n}\rangle \langle{\Omega_n},{\Omega_n}\rangle^{-1} \langle{\Omega_n},{\Psi_n}\rangle \\
&\quad=\langle{\Psi_n^c},{\Omega_n}\rangle \langle{\Omega_n},{\Omega_n}\rangle^{-1} \langle{\Omega_n},{\Psi_n^c}\rangle \\
&\qquad+\langle{\Psi_n^a},{\Omega_n}\rangle \langle{\Omega_n},{\Omega_n}\rangle^{-1} \langle{\Omega_n},{\Psi_n^c}\rangle \\
&\qquad+\langle{\Psi_n^c},{\Omega_n}\rangle \langle{\Omega_n},{\Omega_n}\rangle^{-1} \langle{\Omega_n},{\Psi_n^a}\rangle \\
&\qquad+\langle{\Psi_n^a},{\Omega_n}\rangle \langle{\Omega_n},{\Omega_n}\rangle^{-1} \langle{\Omega_n},{\Psi_n^a}\rangle \\
&\quad= \langle{\Psi_n^c},{\Omega_n}\rangle \langle{\Omega_n},{\Omega_n}\rangle^{-1} \langle{\Omega_n},{\Psi_n^c}\rangle
\end{aligned}
\label{eq:Psiconly}
\end{equation}
because $\langle{\Psi_n^a},{\Omega_n}\rangle=0=\langle{\Omega_n},{\Psi_n^a}\rangle$ $\forall n$, as ${\Omega_n}$ is causal and ${\Psi_n^a}$ is anti-causal.

Now, we will construct an approximation of ${\Psi_n^c}$ whose rows can be written using a linear combination of the rows of $\Omega_n$.
To facilitate this, we define (arguments are omitted for notational simplicity)
\begin{equation}
\tilde{\Omega}_n := 
\begin{bmatrix}
\Gamma_n F_1 F_4 & \Gamma_n F_3 F_4 \\
\Gamma_n (F_3 - F_1 F_4) & 0
\end{bmatrix},
\end{equation}
which has been obtained by multiplying the first $n$ rows of $\Omega_n$ by $F_4$, and subtracting the newly obtained first $n$ rows from the last $n$ rows.
Then, if it is possible to find a linear combination of the rows of $\tilde{\Omega}_n$ to write $\Psi^c_n$, the same is possible for $\Omega_n$.

Let $\bar{F}_3:=F_3 - F_1 F_4$, whose inverse is exponentially stable by assumption.
In addition, let row $i$ of $\bar{F}_3^{-1}{\Psi_n^c}$ be given by
\begin{equation}
\bar{F}_3^{-1}{\Psi_n^{c}}(i) =:
\begin{bmatrix}
\sum_{k=1}^\infty \beta^{(i)}_k q^{-k} & 0
\end{bmatrix} ,
\end{equation}
where $|\beta^{(i)}_k|\leq C\lambda^k$ with $\lambda<1$, and
\begin{equation}
{\hat{\Psi}_n^{c}}(i) :=
\begin{bmatrix}
\sum_{k=1}^n \beta^{(i)}_k \bar F_3 q^{-k} & 0
\end{bmatrix} 
\end{equation}
be the row $i$ of a matrix ${\hat{\Psi}_n^{c}}(i)$.
We re-write the right side of~\eqref{eq:Psiconly} as
\begin{equation}
\langle{\Psi_n^c},{\Omega_n}\rangle \langle{\Omega_n},{\Omega_n}\rangle^{-1} \langle{\Omega_n},{\Psi_n^c}\rangle = \Lambda_n+\Delta^{(1)}_n,
\label{eq:Lambda+Delta}
\end{equation}
where
\begin{equation}
\begin{aligned}
\Lambda_n = &\langle\hat{\Psi}_n^c,{\Omega_n}\rangle \langle{\Omega_n},{\Omega_n}\rangle^{-1} \langle{\Omega_n},\hat{\Psi}_n^c\rangle \\
\Delta^{(1)}_n = &\langle{\Psi_n^c}-\hat{\Psi}_n^c,{\Omega_n}\rangle \langle{\Omega_n},{\Omega_n}\rangle^{-1} \langle{\Omega_n},{\Psi_n^c}\rangle \\
&+ \langle{\Psi_n^c},{\Omega_n}\rangle \langle{\Omega_n},{\Omega_n}\rangle^{-1} \langle{\Omega_n},{\Psi_n^c}-\hat{\Psi}_n^c\rangle \\ 
&+ \langle{\Psi_n^c}-\hat{\Psi}_n^c,{\Omega_n}\rangle \langle{\Omega_n},{\Omega_n}\rangle^{-1} \langle{\Omega_n},{\Psi_n^c}-\hat{\Psi}_n^c\rangle
\end{aligned}
\label{eq:Lambda_and_Delta}
\end{equation}
By construction, ${\hat{\Psi}_n^{c}}(i)$ is a linear combination of the rows of $\tilde{\Omega}_n$ (and hence, of $\Omega_n$); therefore, ${\hat{\Psi}_n^{c}}\in\mathcal{S}_{\Omega_n}$ and
\begin{equation}
\Lambda_n = \langle\hat{\Psi}_n^c,\hat{\Psi}_n^c\rangle.
\label{eq:Lambda}
\end{equation}
Using~\eqref{eq:Psiconly}, \eqref{eq:Lambda+Delta}, and~\eqref{eq:Lambda}, we write
\begin{equation}
\langle{\Psi_n},{\Omega_n}\rangle \langle{\Omega_n},{\Omega_n}\rangle^{-1} \langle{\Omega_n},{\Psi_n}\rangle = \langle\hat{\Psi}_n^c,\hat{\Psi}_n^c\rangle + \Delta^{(1)}_n.
\label{eq:innerproj_almostend}
\end{equation}
Moreover, we can re-write~\eqref{eq:innerproj_almostend} as
\begin{equation}
\langle{\Psi_n},{\Omega_n}\rangle \langle{\Omega_n},{\Omega_n}\rangle^{-1} \langle{\Omega_n},{\Psi_n}\rangle = \langle{\Psi_n^c},{\Psi_n^c}\rangle + \Delta_n,
\label{eq:innerproj_end}
\end{equation}
where $\Delta_n = \Delta^{(1)}_n + \Delta^{(2)}_n$ and
\begin{equation}
\Delta^{(2)}_n = \langle{\Psi_n^c}-\hat{\Psi}_n^c,{\Psi_n^c}\rangle + \langle{\Psi_n^c},{\Psi_n^c}-\hat{\Psi}_n^c\rangle + \langle{\Psi_n^c}-\hat{\Psi}_n^c,{\Psi_n^c}-\hat{\Psi}_n^c\rangle .
\end{equation}
Replacing~\eqref{eq:innerproj_end} in~\eqref{eq:proj}, we obtain
\begin{multline}
\lim\limits_{n\to\infty} \langle\gamma,{\Psi_n}\rangle [ \langle{\Psi_n},{\Omega_n}\rangle \langle{\Omega_n},{\Omega_n}\rangle^{-1} \langle{\Omega_n},{\Psi_n}\rangle ]^{-1} \langle{\Psi_n},\gamma\rangle \\ = \lim\limits_{n\to\infty} \langle\gamma,{\Psi_n}\rangle [ \langle{\Psi_n^c},{\Psi_n^c}\rangle + \Delta_n ]^{-1} \langle{\Psi_n},\gamma\rangle ,
\label{eq:outerproj}
\end{multline}
and using the Sherman-Morrison-Woodbury formula, we re-write~\eqref{eq:outerproj} as
\begin{equation}
\begin{aligned}
&\lim\limits_{n\to\infty} \langle\gamma,{\Psi_n}\rangle [ \langle{\Psi_n^c},{\Psi_n^c}\rangle + \Delta_n ]^{-1} \langle{\Psi_n},\gamma\rangle \\
&= \lim\limits_{n\to\infty} \langle\gamma,{\Psi_n}\rangle \langle{\Psi_n^c},{\Psi_n^c}\rangle^{-1} \langle{\Psi_n},\gamma\rangle \\
&\quad+ \lim\limits_{n\to\infty} \langle\gamma,{\Psi_n}\rangle \langle{\Psi_n^c},{\Psi_n^c}\rangle^{-1} \Delta_n \\ 
&\qquad\cdot[I+\langle{\Psi_n^c},{\Psi_n^c}\rangle^{-1}\Delta_n]^{-1} \langle{\Psi_n^c},{\Psi_n^c}\rangle^{-1} \langle{\Psi_n},\gamma\rangle.
\end{aligned}
\label{eq:matrixinvlemma}
\end{equation}

We want to show that the second term on the right-hand side of~\eqref{eq:matrixinvlemma}, for which we can write
\begin{equation}
\begin{aligned}
&\begin{multlined}
||\langle\gamma,{\Psi_n}\rangle \langle{\Psi_n^c},{\Psi_n^c}\rangle^{-1} \Delta_n [I+\langle{\Psi_n^c},{\Psi_n^c}\rangle\Delta_n]^{-1} \\ \cdot \langle{\Psi_n^c},{\Psi_n^c}\rangle^{-1} \langle{\Psi_n},\gamma\rangle||
\end{multlined} \\ 
&\begin{multlined}[.8\displaywidth]
\quad \leq ||\langle\gamma,{\Psi_n}\rangle||^2 \; ||\langle{\Psi_n^c},{\Psi_n^c}\rangle^{-1}||^2 \; \\ \cdot ||[I+\langle{\Psi_n^c},{\Psi_n^c}\rangle^{-1}\Delta_n]^{-1}|| \; ||\Delta_n||,
\end{multlined}
\end{aligned}
\label{eq:normaftermatrixinv}
\end{equation}
tends to zero as $n$ tends to infinity.
We start by considering the term $\Delta_n$, for which we will need that
\begin{equation}
\langle{\Psi_n^c},{\Psi_n^c}\rangle =
\frac{1}{2\pi} \int_{-\pi}^{\pi} P_n \Gamma_n \Gamma_n^* P_n^\prime d\omega = P_n P_n^\prime.
\label{eq:PPt}
\end{equation}
Using also~\eqref{eq:CS-matrices} and~\eqref{eq:H2bound}, we can write
\begin{multline}
||\Delta_n|| \leq 2 ||\langle{\Omega_n},{\Omega_n}\rangle|| \; ||\langle{\Omega_n},{\Omega_n}\rangle^{-1}|| \; ||P_n|| \; ||\Psi_n^c-\hat{\Psi}_n^c||_{\mathcal{H}_2} \\
+ ||\langle{\Omega_n},{\Omega_n}\rangle||  \; ||\langle{\Omega_n},{\Omega_n}\rangle^{-1}|| \; ||\Psi_n^c-\hat{\Psi}_n^c||_{\mathcal{H}_2}^2 \\
+ 2 ||P_n|| \; ||\Psi_n^c-\hat{\Psi}_n^c||_{\mathcal{H}_2} + ||\Psi_n^c-\hat{\Psi}_n^c||_{\mathcal{H}_2}^2 .
\label{eq:Deltaineq}
\end{multline}
For row $i$ of ${\Psi_n^c} - \hat{\Psi}_n^c$, we have that 
\begin{equation}
\begin{aligned}
||{\Psi_n^{c,i}} - {\hat{\Psi}_n^{c,i}}|| &= \textstyle || \bar F_3 \sum_{k=1}^\infty \beta^{(i)}_k q^{-k} - \sum_{k=1}^n \beta^{(i)}_k \bar F_3 q^{-k} || \\
&\leq ||\bar F_3|| \, || \textstyle \sum_{k=n+1}^\infty \beta^{(i)}_k q^{-k} || \leq C \lambda^n 
\end{aligned}
\end{equation}
Then, 
\begin{equation}
||\Psi_n^c-\hat{\Psi}_n^c||_{\mathcal{H}_2} \leq C \sqrt{n} \lambda^n \to 0 \text{ as } n\to\infty.
\label{eq:Psidiff}
\end{equation}
Moreover, using~\eqref{eq:CS-matrices}, we have
\begin{equation}
||P_n|| = \left|\left| \frac{1}{2\pi} \int_{-\pi}^{\pi} \Gamma_n \Gamma_n^* F_5^* \text{d}\omega \right|\right| \leq ||F_5^*||_{\mathcal{H}_\infty} \leq C,
\label{eq:Pnbound}
\end{equation}
and, using~\eqref{eq:ToepBound} and
\begin{equation}
\begin{multlined}
\langle{\Omega_n},{\Omega_n}\rangle
=
\frac{1}{2\pi} \int_{-\pi}^{\pi} \\
\!\!\!\!\!\!\begin{bmatrix}
\Gamma_n\Gamma_n^* (|F_1|^2 + |F_2|^2) & \Gamma_n\Gamma_n^* (F_1F_3^* + |F_2|^2 F_4^*)\\
\Gamma_n\Gamma_n^* (F_3F_1^*+|F_2|^2F_4) & \Gamma_n\Gamma_n^* (|F_3|^2+|F_2 F_4|^2)
\end{bmatrix} \text{d} \omega,
\end{multlined}
\end{equation}
we have
\begin{equation}
\begin{aligned}
||\langle{\Omega_n},{\Omega_n}\rangle||
\leq &||\,|F_1|^2 + |F_2|^2||_{\mathcal{H}_\infty} \\
&+ 2 ||F_1F_3^*+|F_2|^2F_4^*||_{\mathcal{H}_\infty} \\
&+ ||\,|F_3|^2+|F_2 F_4|^2||_{\mathcal{H}_\infty} \\
\leq &C.
\end{aligned}
\label{eq:Omegabound}
\end{equation}
By assumption, it follows from $\sigma_\text{min}(\langle{\Omega_n},{\Omega_n}\rangle)<C$ that $||\langle{\Omega_n},{\Omega_n}\rangle^{-1}||<C$.
Together with~\eqref{eq:Deltaineq}, \eqref{eq:Psidiff}, and~\eqref{eq:Omegabound}, we have
\begin{equation}
||\Delta_n|| \to 0 \text{ as } n\to \infty.
\label{eq:Deltan->0}
\end{equation}
Then, if the remaining matrix norms in the right-hand side of~\eqref{eq:normaftermatrixinv} are bounded, this term will tend to zero as $n\to\infty$.
For $\langle\gamma,\Psi_n\rangle$, we have
\begin{equation}
\begin{aligned}
||\langle\gamma,\Psi_n\rangle|| &= \left|\left|
\frac{1}{2\pi} \int_{-\pi}^{\pi}
\begin{bmatrix}
\Gamma_n\Gamma_n^* F_5 F_6  \\
\Gamma_n\Gamma_n^* F_5 F_7 
\end{bmatrix} \text{d} \omega \right|\right| \\
&\leq ||F_5F_6||_{\mathcal{H}_\infty} + ||F_5F_7||_{\mathcal{H}_\infty} \leq C.
\end{aligned}
\label{eq:norm_deltaPsi}
\end{equation}
Also, we have that
\begin{equation}
\begin{aligned}
||\langle{\Psi_n^c},{\Psi_n^c}\rangle^{-1}&|| = P_n^{-\prime} P_n^{-1} \\ &= \frac{1}{2\pi} \int_{-\pi}^\pi \Gamma_n \Gamma_n^* F_5^{-1} \text{d}\omega \; \frac{1}{2\pi} \int_{-\pi}^\pi \Gamma_n \Gamma_n^* F_5^{-*} \text{d}\omega \\ 
&\leq ||F_5||^2_{\mathcal{H}_\infty} \leq C .
\end{aligned}
\label{eq:invPsicPsic}
\end{equation}
Together with~\eqref{eq:matrixinvlemma}, \eqref{eq:normaftermatrixinv}, \eqref{eq:Deltan->0}, and \eqref{eq:norm_deltaPsi}, we have
\begin{multline}
\lim\limits_{n\to\infty} \langle\gamma,{\Psi_n}\rangle [ \langle{\Psi_n},{\Omega_n}\rangle \langle{\Omega_n},{\Omega_n}\rangle^{-1} \langle{\Omega_n},{\Psi_n}\rangle ]^{-1} \langle{\Psi_n},\gamma\rangle \\ = \lim\limits_{n\to\infty} \langle\gamma,{\Psi_n}\rangle \langle{\Psi_n^c},{\Psi_n^c}\rangle^{-1} \langle{\Psi_n},\gamma\rangle.
\label{eq:firstproj_done}
\end{multline}

\subsubsection*{Outer projection: $\langle\gamma,{\Psi_n}\rangle \langle{\Psi_n^c},{\Psi_n^c}\rangle^{-1} \langle{\Psi_n},\gamma\rangle$}

Recalling that ${\Psi_n}$ can be written as~\eqref{eq:Psic+a}, we use that $\gamma$ is causal and ${\Psi_n^a}$ anti-causal to conclude that, analogously to~\eqref{eq:Psiconly},
\begin{multline}
\lim\limits_{n\to\infty} \langle\gamma,{\Psi_n}\rangle \langle{\Psi_n^c},{\Psi_n^c}\rangle^{-1} \langle{\Psi_n},\gamma\rangle \\ = \lim\limits_{n\to\infty} \langle\gamma,{\Psi_n^c}\rangle \langle{\Psi_n^c},{\Psi_n^c}\rangle^{-1} \langle{\Psi_n^c},\gamma\rangle.
\label{eq:Psiconly_2}
\end{multline}
Now, row $i$ of $\gamma$ can be written as
\begin{equation}
\gamma(i) =:
\begin{bmatrix}
\sum_{k=1}^\infty s_k^{(i)} q^{-k} & \;\;0
\end{bmatrix},
\end{equation}
where
$|s_k^{(i)}|\leq C\lambda^k$, $\lambda<1$
due to exponential stability.
Let also
$\gamma^{n}(i) :=
[\;\sum_{k=1}^n s_k^{(i)} q^{-k} \;\;\; 0 \;]$
be row $i$ of a matrix $\gamma^n$.
We re-write the right side of~\eqref{eq:Psiconly_2} as
\begin{multline}
\lim\limits_{n\to\infty} \langle\gamma,{\Psi_n^c}\rangle \langle{\Psi_n^c},{\Psi_n^c}\rangle^{-1} \langle{\Psi_n^c},\gamma\rangle \\= \lim\limits_{n\to\infty} \langle\gamma^n,{\Psi_n^c}\rangle \langle{\Psi_n^c},{\Psi_n^c}\rangle^{-1} \langle{\Psi_n^c},\gamma^n\rangle + \lim\limits_{n\to\infty} \Delta^{(3)}_n,
\label{eq:lastproj_aux}
\end{multline}
where
\begin{multline}
\Delta^{(3)}_n = \langle\gamma-\gamma^n,{\Psi_n^c}\rangle \langle{\Psi_n^c},{\Psi_n^c}\rangle^{-1} \langle{\Psi_n^c},\gamma\rangle \\ + \langle\gamma,{\Psi_n^c}\rangle \langle{\Psi_n^c},{\Psi_n^c}\rangle^{-1} \langle{\Psi_n^c},\gamma-\gamma^n\rangle \\ + \langle\gamma-\gamma^n,{\Psi_n^c}\rangle \langle{\Psi_n^c},{\Psi_n^c}\rangle^{-1} \langle{\Psi_n^c},\gamma-\gamma^n\rangle .
\end{multline}
Using a similar approach for $\Delta^{(3)}_n$ as we did for $\Delta_n$, it can be shown that
\begin{equation}
||\Delta_n^{(3)}|| \leq C ( ||\gamma-\gamma^n||_{\mathcal{H}_2} + ||\gamma-\gamma^n||_{\mathcal{H}_2}^2 ) \to 0 \text{ as } n\to\infty.
\label{eq:Delta3_O}
\end{equation}
Thus, \eqref{eq:lastproj_aux} reduces to
\begin{multline}
\lim\limits_{n\to\infty} \langle\gamma,{\Psi_n^c}\rangle \langle{\Psi_n^c},{\Psi_n^c}\rangle^{-1} \langle{\Psi_n^c},\gamma\rangle \\= \lim\limits_{n\to\infty} \langle\gamma^n,{\Psi_n^c}\rangle \langle{\Psi_n^c},{\Psi_n^c}\rangle^{-1} \langle{\Psi_n^c},\gamma^n\rangle .
\label{eq:lastproj}
\end{multline}

Finally, we have that
\begin{equation}
\begin{aligned}
\langle\gamma^n,{\Psi_n^c}\rangle \langle{\Psi_n^c},{\Psi_n^c}\rangle^{-1} \langle{\Psi_n^c},\gamma^n\rangle &= \langle\text{Proj}_{\mathcal{S}_{{\Psi_n^c}}}\gamma^n,\text{Proj}_{\mathcal{S}_{{\Psi_n^c}}}\gamma^n\rangle \\ &= \langle \gamma^n, \gamma^n \rangle ,
\end{aligned}
\label{eq:lastproj_solved}
\end{equation}
where the last equality follows from $\gamma^n \in \mathcal{S}_{{\Psi_n^c}}$,
as consequence of $P_n$ being invertible.
Thus, replacing~\eqref{eq:lastproj_solved} in~\eqref{eq:lastproj}, we have
\begin{equation}
\lim\limits_{n\to\infty} \langle\gamma,{\Psi_n^c}\rangle \langle{\Psi_n^c},{\Psi_n^c}\rangle^{-1} \langle{\Psi_n^c},\gamma\rangle = \lim\limits_{n\to\infty} \langle\gamma^n,\gamma^n\rangle = \langle\gamma,\gamma\rangle,
\end{equation}
which, together with~\eqref{eq:firstproj_done}, \eqref{eq:Psiconly_2}, \eqref{eq:lastproj_aux}, and~\eqref{eq:Delta3_O} implies~\eqref{eq:proj}, as we wanted to show. \hfill\hfill \qed

\section{Proof of Theorem~\ref{thm:consistencyWLS}}
\label{sec:theorem2_proof}

We will show that
\begin{equation}
||\hat{\theta}_N^\text{WLS} - \theta_\nul||\to 0, \wpone .
\label{eq:thetawls-theta0_norm->0}
\end{equation}
To do this, we use~\eqref{eq:thetawls-theta0_expand} to write
\begin{multline}
\norm{\thetawls-\theta_\nul} \leq \norm{M^{-1}(\hat{\eta}_N,\thetals)} \norm{Q_n(\hat{\eta}_N)} \\ \cdot \norm{W_n(\thetals)} \norm{T_n(\theta_\nul)} \norm{\hat{\eta}_N-\eta^{n(N)}_\nul} .
\label{eq:thetawls-theta0_leq}
\end{multline}
From \cite{wnsf_tac}, we have that
\begin{gather}
||\hat{\eta}_N-\eta^{n(N)}_\nul|| \to 0 \wpone,
\label{eq:etahat-eta0->0} \\
||T_n(\theta_\nul)|| \leq C \quad \forall n ,
\label{eq:T0bound}
\end{gather}
and that, w.p.1, there exists $\bar{N}$ such that
\begin{equation}
||Q_n(\hat{\eta}_N)|| \leq C \quad \forall N>\bar{N} .
\label{eq:Qnhat_bounded}
\end{equation}
Then, we have left to show that $W_n(\thetals)$ is bounded and $M(\hat{\eta}_N,\thetals)$ is invertible for sufficiently large $N$.

We begin by considering the inverse of $W_n(\thetals)$, for which we have
\begin{equation}
\begin{multlined}
||T_n(\thetals)[R^n_N]^{-1}T_n^\prime(\thetals)|| \leq ||T_n(\theta_\nul)[\bar{R}^n]^{-1}T_n^\prime(\theta_\nul)|| \\ + ||T_n(\thetals)[R^n_N]^{-1}T_n^\prime(\thetals)-T_n(\theta_\nul)[\bar{R}^n]^{-1}T_n^\prime(\theta_\nul)||.
\end{multlined}
\label{eq:invW_norm}
\end{equation}
In turn, it can be shown that
\begin{equation}
T_n(\theta_\nul)[\bar{R}^n]^{-1}T_n^\prime(\theta_\nul) = \langle{\Psi_n},{\Omega_n}\rangle \langle{\Omega_n},{\Omega_n}\rangle^{-1} \langle{\Omega_n},{\Psi_n}\rangle ,
\end{equation}
where ${\Psi_n}$ and ${\Omega_n}$ are given by~\eqref{eq:gammaPsiOmega_original}, which satisfy the conditions of Theorem~\ref{thm:geometric} with 
\begin{equation}
\begin{array}{cc}
F_1(q)=-G_\nul(q)S_\nul(q) F_r(q), \;\; & \;\; F_2(q)= -H_\nul(q) S_\nul(q)\sigma_\nul, \\
F_3(q)=S_\nul(q)F_r(q), \;\; & \;\; F_4(q) = K(q) .
\end{array}
\end{equation}
In particular, there is $\bar{n}$ such that $\langle{\Omega_n},{\Omega_n}\rangle=\bar{R}^n$ is invertible for all $n>\bar{n}$~\citep{ljung&wahlberg92}, and $F_3(q)-F_1(q)F_4(q)=F_r(q)$ can be chosen to have a stable inverse under the constraint $\Phi_r(e^{i\omega})=|F_r(e^{i\omega})|^2$.
From~\eqref{eq:innerproj_end}, \eqref{eq:PPt}, \eqref{eq:Pnbound}, and~\eqref{eq:Deltan->0}, we have that there is $\bar{n}$ such that
\begin{equation}
||T_n(\theta_\nul)[\bar{R}^n]^{-1}T_n^\prime(\theta_\nul)|| = ||\langle{\Psi_n^c},{\Psi_n^c}\rangle + \Delta_n|| \leq C \; \forall n>\bar{n}.
\label{eq:Winv0bounded}
\end{equation}
Then, and using also~\eqref{eq:invPsicPsic}, we have
\begin{multline}
||\bar{W}(\theta_\nul)|| := ||[T_n(\theta_\nul)[\bar{R}^n]^{-1}T_n^\prime(\theta_\nul)]^{-1}|| \\ = ||(\langle{\Psi_n^c},{\Psi_n^c}\rangle+\Delta_n)^{-1}|| \leq C \quad \forall n>\bar{n}.
\label{eq:W0bounded}
\end{multline}
Concerning the second term on the right-hand side of~\eqref{eq:invW_norm}, we can write
\begin{equation}
\begin{aligned}
&||T_n(\thetals)[R^n_N]^{-1}T_n^\prime(\thetals)-T_n(\theta_\nul)[\bar{R}^n]^{-1}T_n^\prime(\theta_\nul)|| \\
&\leq ||T_n(\thetals)-T_n(\theta_\nul)|| \; ||[R^n_N]^{-1}|| \; ||T_n(\thetals)|| \\
&\quad+ ||T_n(\thetals)-T_n(\theta_\nul)|| \; ||[R^n_N]^{-1}|| \; ||T_n(\theta_\nul)|| \\
&\quad+ ||T_n(\theta_\nul)||^2 \; ||[R^n_N]^{-1}-[\bar{R}^n]^{-1}||
\end{aligned}
\label{eq:Winvhat-Winv0}
\end{equation}
Now, the results by \cite{wnsf_tac} apply to~\eqref{eq:Winvhat-Winv0}.
In particular, there is $\bar{N}$ such that
\begin{equation}
\begin{aligned}
||[R^n_N]^{-1}|| \leq C \quad \forall N>\bar{N}, &\qquad 
||[R^n_N]|| \leq C \quad \forall N>\bar{N}, \\
||T_n(\thetals)|| \leq C &\quad \forall N>\bar{N} 
\end{aligned}
\label{eq:Tls_RnN_bounded}
\end{equation}
w.p.1, and that
\begin{gather}
||T_n(\thetals)-T_n(\theta_\nul)|| \to 0 \wpone , \\
\begin{multlined}
||[R^n_N]^{-1}-[\bar{R}^n]^{-1}||\leq ||\bar{R}^n_N|| \; ||\bar{R}^n-R^n_N|| \; ||[R^n_N]^{-1}|| \\ \to 0 \wpone .
\end{multlined}
\end{gather}
Together with~\eqref{eq:T0bound}, we conclude that
\begin{multline}
||T_n(\thetals)[R^n_N]^{-1}T_n^\prime(\thetals)-T_n(\theta_\nul)[\bar{R}^n]^{-1}T_n^\prime(\theta_\nul)|| \\
\to 0 \wpone .
\label{eq:Winvhat-Winv0->0}
\end{multline}
Using~\eqref{eq:Winvhat-Winv0->0}, \eqref{eq:Winv0bounded}, and~\eqref{eq:invW_norm}, there is $\bar{N}$ such that
\begin{equation}
||T_n(\thetals)[R^n_N]^{-1}T_n^\prime(\thetals)|| \leq C \quad \forall N>\bar{N} \quad \text{w.p.1.}
\end{equation}
Because of~\eqref{eq:Winvhat-Winv0->0} and invertibility of $T_n(\theta_\nul)[\bar{R}^n]^{-1}T_n^\prime(\theta_\nul)$, by continuity of eigenvalues there is $\bar{N}$ such that $W(\thetals)=[T_n(\thetals)[R^n_N]^{-1}T_n^\prime(\thetals)]^{-1}$ exists for all $N>\bar{N}$, and
\begin{equation}
||W(\thetals)-\bar{W}(\theta_\nul)|| \to 0, \wpone .
\label{eq:What-W0->0}
\end{equation}
Moreover, \eqref{eq:What-W0->0} and~\eqref{eq:W0bounded} imply that, w.p.1,
\begin{equation}
||W(\thetals)||\leq C \quad \forall N>\bar{N} .
\label{eq:What_bounded}
\end{equation}

Having shown~\eqref{eq:What_bounded}, we have left to show that $M(\hat{\eta}_N,\thetals)$ is invertible for sufficiently large $N$, in order to show~\eqref{eq:thetawls-theta0_norm->0}.
Recall the definition~\eqref{eq:Mbar}, which can alternatively be written as~\eqref{eq:Mbar_proj}, where $\gamma$ is given by~\eqref{eq:gammaPsiOmega_original}.
Then, from Theorem~\ref{thm:geometric} with
\begin{equation}
F_6(q) = -\frac{B_\nul(q)S_\nul(q)F_r(q)}{F_\nul(q)}, \quad F_7(q) = \frac{A_\nul(q)S_\nul(q)F_r(q)}{F_\nul(q)},
\end{equation}
we have that
\begin{equation}
\bar{M}(\eta_\nul,\theta_\nul) = M .
\label{eq:barM=M}
\end{equation}
where $M$ is given by~\eqref{eq:M}. Because the inverse of $M$ corresponds to the CR bound for an open-loop problem with input $u_t=S_\nul(q)r_t$, and the CR bound exists for an informative experiment~\citep{ljung99}, we conclude that $\bar{M}(\eta_\nul,\theta_\nul)$ is invertible.
Then, we analyze the difference
\begin{equation}
\begin{aligned}
&||M(\hat{\eta}_N,\thetals)-Q^\prime_n(\eta^n_\nul) \bar{W}_n(\theta_\nul) Q_n(\eta^n_\nul)|| \\
&\leq ||Q_n(\hat{\eta}_N)-Q_n(\eta^n_N)|| \; ||W(\thetals)|| \; ||Q_n(\hat{\eta}_N)|| \\
&\quad+ ||Q_n(\hat{\eta}_N)-Q_n(\eta^n_N)|| \; ||W(\thetals)|| \; ||Q_n(\eta^n_\nul)|| \\
&\quad+ ||Q_n(\eta^n_N)||^2 \; ||W(\thetals)-\bar{W}(\theta_\nul)||
\end{aligned}
\label{eq:Mhat-M0}
\end{equation}
From \cite{wnsf_tac}, we have that
\begin{gather}
||Q_n(\hat{\eta}_N)-Q_n(\eta^n_N)|| \to 0, \wpone, \\
||Q_n(\eta^n_\nul)|| \leq C \quad \forall n .
\label{eq:Q0bound}
\end{gather}
Together with~\eqref{eq:What-W0->0} and~\eqref{eq:Qnhat_bounded}, we conclude that
\begin{multline}
||M(\hat{\eta}_N,\thetals)-Q^\prime_n(\eta^n_\nul) \bar{W}_n(\theta_\nul) Q_n(\eta^n_\nul)|| \\
\to 0, \wpone.
\label{eq:Mhat-M0->0}
\end{multline}
Using~\eqref{eq:Mhat-M0->0}, invertibility of $\bar{M}(\eta_\nul,\theta_\nul)$, and continuity of eigenvalues, we have that there is $\bar{N}$ such that $M(\hat{\eta}_N,\thetals)$ is invertible for all $N>\bar{N}$,
\begin{equation}
||M^{-1}(\hat{\eta}_N,\thetals)|| \leq C \quad \forall N>\bar{N},
\label{eq:Mhatinv_bounded}
\end{equation}
and, using also~\eqref{eq:barM=M} and~\eqref{eq:Mbar},
\begin{equation}
M^{-1}(\hat{\eta}_N,\thetals) \to M^{-1} \wpone.
\label{eq:invMhat-invM0->0}
\end{equation}

Finally, using~\eqref{eq:Mhatinv_bounded}, \eqref{eq:What_bounded}, \eqref{eq:Qnhat_bounded}, \eqref{eq:T0bound}, \eqref{eq:etahat-eta0->0}, and~\eqref{eq:thetawls-theta0_leq}, we conclude that~\eqref{eq:thetawls-theta0_norm->0} is satisfied, as we wanted to show. \hfill\hfill \qed

\section{Proof of Theorem~\ref{thm:asycov}}
\label{sec:theorem3_proof}

In this appendix, we prove Theorem~\ref{thm:asycov}.
We begin by reformulating~\eqref{eq:thetawls-theta0_expand} as
\begin{equation}
\sqrt{N} (\thetawls-\theta_\nul) = M^{-1}(\hat{\eta}_N,\thetals) x(\theta_\nul;\hat{\eta}_N,\thetals) ,
\end{equation}
where 
\begin{equation}
\begin{aligned}
M(\hat{\eta}_N,\thetals) &= Q_n^\prime(\hat{\eta}_N) W_n(\thetals) Q_n(\hat{\eta}_N) , \\
x(\theta_\nul;\hat{\eta}_N,\thetals) &= \sqrt{N} Q_n^\prime(\hat{\eta}_N) W_n(\thetals) T_n(\theta_\nul) (\hat{\eta}_N-\eta^{n(N)}_\nul) .
\end{aligned}
\label{eq:Mx}
\end{equation}
If we assume that
\begin{equation}
x(\theta_\nul;\hat{\eta}_N,\thetals) \sim As \mathcal{N} (0,X) ,
\label{eq:xdist}
\end{equation}
we have that, using~\eqref{eq:invMhat-invM0->0} and Lemma B.4 by~\citet{soderstromstoica89book},
\begin{equation}
\sqrt{N}(\thetawls-\theta_\nul) \sim As\mathcal{N} \left(0,M^{-1} X M^{-1} \right) .
\label{eq:thetawls_dist_MandP}
\end{equation}
We will proceed to show that~\eqref{eq:xdist} is verified with
\begin{equation}
X = \sigma_\nul^2 \lim\limits_{n\to\infty} Q_n^\prime(\eta^n_\nul) \bar{W}_n(\theta_\nul) Q_n(\eta^n_\nul) = \sigma_\nul^2 M,
\label{eq:Px}
\end{equation}
where the second equality follows directly from~\eqref{eq:Mbar} and~\eqref{eq:barM=M}. 
We now proceed to show the first equality.

The idea is, as in \citet{wnsf_tac}, to use Theorem 7.3 by~\citet{ljung&wahlberg92}.
With this purpose, we first show that $x(\theta_\nul;\hat{\eta}_N,\thetals)$ has the same asymptotic distribution and covariance as $\sqrt{N}\Upsilon^n(\hat{\eta}_N-\bar{\eta}^{n(N)})$, where $\Upsilon^n$ is a deterministic matrix.
From~\citet{wnsf_tac}, it follows that $x(\theta_\nul;\hat{\eta}_N,\thetals)$ has the same asymptotic covariance and distribution as
\begin{equation}
\sqrt{N} Q_n^\prime(\eta^{n(N)}_\nul) W_n(\thetals) T_n(\theta_\nul) (\hat{\eta}_N-\bar{\eta}^{n(N)}).
\label{eq:x2}
\end{equation}
The next step is to show that $W_n(\thetals)$ can be replaced by $\bar{W}_n(\theta_\nul)$ in~\eqref{eq:x2} without affecting the asymptotic distribution and covariance.
However, this step does not follow directly from \cite{wnsf_tac}, as inverses of the matrices that compose $W_n$ cannot be taken individually here, because $T_n$ is not square.
In the following, we do this for a non-square $T_n$ matrix.

First, re-write~\eqref{eq:x2} as
\begin{multline}
\sqrt{N} Q_n^\prime(\eta^{n(N)}_\nul) \bar{W}_n(\theta_\nul) T_n(\theta_\nul) (\hat{\eta}_N-\bar{\eta}^{n(N)}) \\
+ \sqrt{N} Q_n^\prime(\eta^{n(N)}_\nul) [W_n(\thetals)-\bar{W}_n(\theta_\nul)] T_n(\theta_\nul) (\hat{\eta}_N-\bar{\eta}^{n(N)}).
\label{eq:x3}
\end{multline}
Then, it follows from Lemma B.4 by~\cite{soderstromstoica89book} that \eqref{eq:x2} and the first term in~\eqref{eq:x3} have the same asymptotic distribution and covariance if the second term in~\eqref{eq:x3} tends to 0 as $N\to\infty$ w.p.1.
We proceed to show this.
Consider
\begin{equation}
\begin{aligned}
&||\sqrt{N} Q_n^\prime(\eta^{n(N)}_\nul) [W_n(\thetals)-\bar{W}_n(\theta_\nul)] T_n(\theta_\nul) (\hat{\eta}_N-\bar{\eta}^{n(N)})|| \\ 
&\begin{multlined}[.8\displaywidth]
\leq \sqrt{N} ||Q_n^\prime(\eta^{n(N)}_\nul)|| \; ||W_n(\thetals)-\bar{W}_n(\theta_\nul)|| \\ \cdot ||T_n(\theta_\nul)|| \; ||\hat{\eta}_N-\bar{\eta}^{n(N)}|| 
\end{multlined}\\
&\leq C \sqrt{N} ||W_n(\thetals)-\bar{W}_n(\theta_\nul)|| \;||\hat{\eta}_N-\bar{\eta}^{n(N)}|| ,
\end{aligned}
\label{eq:x4}
\end{equation}
where the last inequality follows from~\eqref{eq:T0bound} and~\eqref{eq:Q0bound}.
Writing
\begin{equation}
W_n(\thetals)-\bar{W}_n(\theta_\nul) = \bar{W}_n(\theta_\nul) [\bar{W}^{-1}_n(\theta_\nul)-W^{-1}_n(\thetals)] W_n(\thetals),
\end{equation}
and because~\eqref{eq:What_bounded} and~\eqref{eq:W0bounded} guarantee that $\bar{W}_n(\theta_\nul)$ and $W_n(\thetals)$ are bounded (in the latter, for sufficiently large $N$), it follows from~\eqref{eq:x4} that
\begin{equation}
\begin{aligned}
||\sqrt{N} &Q_n^\prime(\eta^{n(N)}_\nul) [W_n(\thetals)\!-\!\bar{W}_n(\theta_\nul)] T_n(\theta_\nul) (\hat{\eta}_N-\bar{\eta}^{n(N)})|| \\ 
&\leq C \sqrt{N} ||W_n^{-1}(\thetals)-\bar{W}_n^{-1}(\theta_\nul)|| \;||\hat{\eta}_N-\bar{\eta}^{n(N)}||.
\end{aligned}
\end{equation}
Now, the term $||W_n^{-1}(\thetals)-\bar{W}_n^{-1}(\theta_\nul)||$ corresponds to~\eqref{eq:Winvhat-Winv0}; so, using~\citep{wnsf_tac}
\begin{equation}
||T_n(\thetals)-T_n(\theta_\nul)|| = \mathcal{O} \left( \sqrt{n^2(N)\frac{\log N}{N}} (1+d(N)) \right),
\label{eq:Tnls-Tn0_O}
\end{equation}
\eqref{eq:T0bound} and~\eqref{eq:Tls_RnN_bounded}, the first two terms on the right-hand side of~\eqref{eq:Winvhat-Winv0} decay with~\eqref{eq:Tnls-Tn0_O}.
For the third term, we first write
\begin{equation}
||[R^n_N]^{-1}-[\bar{R}^n]^{-1}|| \leq ||\bar{R}^n_N|| \; ||\bar{R}^n-R^n_N|| \; ||[R^n_N]^{-1}||.
\label{eq:RnN-barR_expand}
\end{equation}
\citet{ljung&wahlberg92} show that
\begin{equation}
||\bar{R}^n-R^n_N|| = \mathcal{O} \left( 2\sqrt{n^2(N)\frac{\log N}{N}} + C\frac{n^2(N)}{N}\right);
\label{eq:barR-RnN_O}
\end{equation}
then, using also~\eqref{eq:RnN-barR_expand}, \eqref{eq:Tls_RnN_bounded}, and~\eqref{eq:T0bound}, we have that the third term on the right-hand side of~\eqref{eq:Winvhat-Winv0} decays according to~\eqref{eq:barR-RnN_O}.
Thus, we have that
\begin{equation}
||W_n(\thetals)-\bar{W}_n(\theta_\nul)|| = \mathcal{O} \left( 2\sqrt{n^2(N)\frac{\log N}{N}} \right),
\label{eq:Wnls-barWn0_O}
\end{equation}
considering only the slowest-decaying term.
Then, from~\eqref{eq:x4}, \eqref{eq:Wnls-barWn0_O}, and~\citep{ljung&wahlberg92}
\begin{equation}
||\hat{\eta}_N-\bar{\eta}^{n(N)}|| = \mathcal{O}\left( \sqrt{\frac{n(N)\log N}{N}} [1+d(N)] \right),
\end{equation}
it follows that
\begin{multline}
C \sqrt{N} ||W_n(\thetals)-\bar{W}_n(\theta_\nul)|| \;||\hat{\eta}_N-\bar{\eta}^{n(N)}|| \to 0 \\ \wpone.
\end{multline}
Finally, using~\eqref{eq:x4} and \eqref{eq:x3}, it follows from Lemma B.4 by~\cite{soderstromstoica89book} that
\begin{equation}
\sqrt{N} Q_n^\prime(\eta^{n(N)}_\nul) \bar{W}_n(\theta_\nul) T_n(\theta_\nul) (\hat{\eta}_N-\bar{\eta}^{n(N)})
\label{eq:x5}
\end{equation}
and~\eqref{eq:x2}---and, in turn, $x(\theta_\nul;\hat{\eta}_N,\thetals)$---have the same asymptotic distribution and covariance.
Thus, we will analyze~\eqref{eq:x5} instead of $x(\theta_\nul;\hat{\eta}_N,\thetals)$.

Applying Theorem 7.3 by~\citet{ljung&wahlberg92} to~\eqref{eq:x5} with $\Upsilon^n=Q_n^\prime(\eta^{n(N)}_\nul) \bar{W}_n(\theta_\nul) T_n(\theta_\nul)$---and recalling that it has the same asymptotic distribution and covariance as $x(\theta_\nul;\hat{\eta}_N,\thetals)$---we have that $x(\theta_\nul;\hat{\eta}_N,\thetals)$ is distributed according to~\eqref{eq:xdist} with
\begin{equation}
\begin{aligned}
&\begin{multlined}[.8\displaywidth]
X = \lim\limits_{n\to\infty} Q_n^\prime(\eta^{n(N)}_\nul) \bar{W}_n(\theta_\nul) T_n(\theta_\nul) \sigma_\nul^2 \\ \cdot \bar{R}^n_N T^\prime_n(\theta_\nul) \bar{W}_n(\theta_\nul) Q_n(\eta^{n(N)}_\nul) \end{multlined}\\
&\;\;\;\;= \sigma_\nul^2 \lim\limits_{n\to\infty} Q_n^\prime(\eta^{n(N)}_\nul) \bar{W}_n(\theta_\nul) Q_n(\eta^{n(N)}_\nul) = \sigma_\nul^2 M,
\end{aligned}
\label{eq:Xfinal}
\end{equation}
where the last equality follows from~\eqref{eq:Mbar}.
Then, replacing~\eqref{eq:Xfinal} in~\eqref{eq:thetawls_dist_MandP}, we obtain
\begin{equation}
\sqrt{N}(\thetawls-\theta_\nul) \sim As\mathcal{N} \left(0,M^{-1} \right) . \qed
\end{equation}

\bibliographystyle{apalike}        
\bibliography{mybib}           

\end{document}